\documentclass[11pt,a4paper]{article}
\pdfoutput=1
\usepackage{jcappub}
\usepackage{subfigure}
\usepackage{mathtools}
\usepackage{multirow}

\def\procspie{\ref@jnl{Proc.~SPIE}}
\def\aap{\ref@jnl{A\&A}}                
\def\apjl{\ref@jnl{ApJ}}                

\newcommand{\be}{\begin{equation}}
\newcommand{\ee}{\end{equation}}
\newcommand{\ba}{\begin{eqnarray}}
\newcommand{\ea}{\end{eqnarray}}
\newcommand{\bi}{\begin{itemize}}
\newcommand{\ei}{\end{itemize}}

\newcommand{\bfi}{\begin{figure}
\epsfxsize=9cm
\epsffile}
\newcommand{\bfinew}{\begin{figure}
\begin{center}
\includegraphics}
\newcommand{\efi}{\end{figure}}
\newcommand{\efinew}{
\end{center}
\end{figure}}
\newcommand{\no}{\nonumber}
\newcommand{\alt}{\lesssim}
\newcommand{\agt}{\gtrsim}

\newcommand{\ang}[1]{\langle #1 \rangle}
\newcommand{\bang}[1]{\Big\langle #1 \Big\rangle}

\graphicspath{ {./figuresnew/} }
\title{Analyzing the cosmic variance limit of remote dipole measurements of the cosmic microwave background using the large-scale kinetic Sunyaev Zel'dovich effect}

\author[a,b]{Alexandra Terrana}
\author[b]{Mary-Jean Harris}
\author[a,b]{Matthew C. Johnson}

\affiliation[a]{Department of Physics and Astronomy, York University, Toronto, Ontario, M3J 1P3, Canada}
\affiliation[b]{Perimeter Institute for Theoretical Physics, Waterloo, Ontario N2L 2Y5, Canada}

\emailAdd{aterrana@perimeterinstitute.ca}
\emailAdd{mharris8@perimeterinstitute.ca}
\emailAdd{mjohnson@perimeterinstitute.ca}

\abstract{Due to cosmic variance we cannot learn any more about large-scale inhomogeneities from the primary cosmic microwave background (CMB) alone. More information on large scales is essential for resolving large angular scale anomalies in the CMB. Here we consider cross correlating the large-scale kinetic Sunyaev Zel'dovich (kSZ) effect and probes of large-scale structure, a technique known as kSZ tomography. The statistically anisotropic component of the cross correlation encodes the CMB dipole as seen by free electrons throughout the observable Universe, providing information about long wavelength inhomogeneities. We compute the large angular scale power asymmetry, constructing the appropriate transfer functions, and estimate the cosmic variance limited signal to noise for a variety of redshift bin configurations. The signal to noise is  significant over a large range of power multipoles and numbers of bins. We present a simple mode counting argument indicating that kSZ tomography can be used to estimate more modes than the primary CMB on comparable scales. A basic forecast indicates that a first detection could be made with next-generation CMB experiments and galaxy surveys. This paper motivates a more systematic investigation of how close to the cosmic variance limit it will be possible to get with future observations.}

\arxivnumber{1610.06919}

\begin{document}
\maketitle
\flushbottom

\section{Introduction}

The cosmic microwave background (CMB) has been an extraordinarily powerful tool for precision cosmology, establishing the standard model, $\Lambda$CDM, at high confidence. However, on very large-scales, CMB measurements are limited by cosmic variance, implying that we can not hope to learn any more about large scale inhomogeneities from the primary CMB alone. Given that nearly all of the hints we have of departures from $\Lambda$CDM are on the very largest scales (for a recent summary of CMB anomalies see~\cite{Schwarz:2015cma}), there is strong motivation to go beyond the primary CMB to learn more. Constraints from probes of Large Scale Structure (LSS), such as next-generation galaxy surveys (e.g~\cite{LSST09}) and 21cm measurements~(e.g.~\cite{2014SPIE.9145E..22B,Abdalla:2015kra}), are poised to become increasingly important for many cosmological parameters. Even still, when it comes to measuring inhomogeneities on scales $\agt {\rm Gpc}$, there will be limited additional constraining power in all but the most ambitious scenarios (e.g. 21cm dark ages cosmology~\cite{Pritchard12}). In this paper, we investigate the viability of an additional probe of large scale inhomogeneities: large scale kinetic Sunyaev Zel'dovich (kSZ) tomography. 

The kSZ effect is a CMB temperature anisotropy arising from the Compton scattering of CMB photons by the bulk motion of free electrons with respect to the CMB rest frame~\cite{SZ80}. There are two main contributions to the kSZ effect: the early-time kSZ effect which arises due to patchy reionization~\cite{Vishniac87,Zhang04a,McQuinn:2005ce} (see e.g.~\cite{Alvarez:2015xzu} for a recent update), and the late-time kSZ effect arising from free electrons in large scale structure (e.g. galaxy clusters) and the intergalactic medium. Thus far, only the latter contribution has been detected. This was achieved by looking for the contribution to CMB temperature anisotropies induced by the pairwise motion of clusters~\cite{Hand12,DeBernardis:2016pdv,Soergel:2016mce,2016A&A...586A.140P}; a (somewhat low significance) detection has also been made at the level of the temperature angular power spectrum~\cite{George14}. Next-generation ``Stage 3" and ``Stage 4" CMB experiments~\cite{Calabrese:2014gwa,Wu:2014hta} will have the ability to make high-significance measurements of the kSZ effect.\footnote{It will at this point become important to separate the early-time and late-time kSZ effects to maximize the science return; for a proposal in this direction see Ref.~\cite{Smith:2016lnt}.} Realizing the full potential of kSZ measurements will continue to rely heavily on cross correlations with future probes of LSS, making the dramatic improvements to come with the next generation of redshift surveys and 21cm measurements equally important. Such cross correlations also open the door to determining the contribution to the global kSZ signal from different redshifts, a technique known as kSZ tomography~\cite{Ho09,Shao11b, Zhang11b, Zhang01,Munshi:2015anr,2016PhRvD..93h2002S,Ferraro:2016ymw,Hill:2016dta}. 

The science case for precision measurements of the kSZ effect is quite broad. In addition to revolutionizing our understanding of reionization, it has the power to probe missing baryons e.g.~\cite{Ho09,2016RAA....16d..15M,2015PhRvL.115s1301H}, make precision tests of gravity~\cite{Xu14,Mueller14,Bianchini:2015iaa}, probe anomalous bulk flows~\cite{Kashlinsky08,Zhang10d,Kashlinsky11,Li12,Lavaux13,Atrio-Barandela14,Planck-I-2014-v}, constrain the properties of dark energy and dark matter~\cite{2015ApJ...808...47M,Xu13}, constrain the masses of neutrinos~\cite{2015PhRvD..92f3501M}, test the Copernican principle~\cite{Goodman95,Zhang11b,Zibin:2014rfa}, constrain the present day vacuum decay rate~\cite{Pen14}, and test the hypothesis that we inhabit an eternally inflating multiverse~\cite{Zhang:2015uta}. 

The contribution to the kSZ effect from each free electron is proportional to the locally observed CMB dipole, and because each free electron probes a different portion of the surface of last scattering, measurements of the kSZ effect can in principle yield information about the homogeneity of the Universe. This is why measurements of the kSZ effect can be so constraining for scenarios that predict a deviation from large-scale homogeneity, such as many of those listed above. Note that this is a dramatically different regime than the one typically explored, for example in the pairwise motion of clusters that yielded the first detection. This large-scale kSZ effect is  sensitive to the Sachs Wolfe and integrated Sachs Wolfe components of the dipole, in addition to the Doppler component from peculiar velocities, and can therefore in principle yield more large-scale information than peculiar velocity surveys or direct measurements of the density field through various tracers of LSS. As mentioned above, cross correlation with tracers of LSS, and therefore kSZ tomography, is key to extracting the most information possible. As we show in more detail below, information about large scale homogeneity is encoded in a statistical anisotropy of the direct correlation of tracers of large scale structure and the small angular scale CMB, e.g. a power asymmetry. Importantly, the contribution to this signal from small-scale peculiar velocities vanishes~\cite{Shao11b}.

The power of kSZ tomography to probe the large scale homogeneity of the Universe has been highlighted previously, notably in refs.~\cite{Zhang10d,Zhang11b,Pen14,Zhang:2015uta}.\footnote{The polarized component of the late-time kSZ effect also has the potential to constrain homogeneity on large scales~\cite{Kamionkowski:1997na,Portsmouth2004,2014PhRvD..90f3518H,Bunn:2006mp,Yasini:2016pby,Liu:2016fqc,Abramo:2006gp}, but we do not consider it further here.} These papers considered theoretical extensions to $\Lambda$CDM where a signal could hopefully be detected with current and near-future experiments. However, as the sensitivity and resolution of CMB experiments continues to develop and as our ability to probe LSS improves, we might hope to enter an era where large scale kSZ tomography becomes a tool not just for constraining exotic scenarios, but for measuring the inhomogeneities we know to exist: those responsible for the large scale temperature anisotropies in the primary CMB. 

The goal of this paper is to explore this eventuality in the most optimistic, cosmic variance limited, scenario. More specifically, we compute the angular spectrum of the asymmetry in the kSZ-LSS cross power expected in $\Lambda$CDM as a function of redshift. Comparing this signal to the accidental power asymmetry expected from the statistically isotropic components of the kSZ effect (the dominant source of CMB temperature anisotropies on small angular scales), we find that the signal-to-noise can be significant ($S/N \sim \mathcal{O} (10^2-10^3)$) over a wide range of angular scales ($\ell_{\rm max} \sim \mathcal{O}(100)$) and in a large number ($N_{\rm bin} \sim \mathcal{O} (10-100)$) of redshift bins. A simple mode counting argument indicates that there is in principle more information in the power asymmetry than in the primary CMB on the relevant scales, for a sufficient number of redshift bins ($N_{\rm bin} \agt 30$). We present a basic forecast, indicating that a first detection could be made with next-generation CMB experiments and galaxy surveys.

The plan of the paper is as follows. In section \ref{sec:ksz}, the large-scale late-time kSZ effect is summarized along with a derivation of the large-scale effective velocity. Section~\ref{sec:simulations} describes simulations of the large-scale effective velocity field. Section \ref{sec:signal} outlines how kSZ tomography can be used to extract the large-scale effective velocity. Then, we derive the cosmic variance limited noise in section~\ref{sec:noise} and estimate the signal using both an RMS estimate and simulations in section~\ref{sec:estsig}. In section~\ref{sec:modes} we provide an estimate for the number of modes that can be obtained using cosmic variance limited kSZ tomography, showing that in principle more information can be extracted than is contained in the primary CMB on comparable scales. Finally, we assess the detectability of the signal with next-generation CMB experiments and galaxy surveys in section~\ref{sec:detectability}; we conclude in section~\ref{sec:conclusion}. A number of results are collected in the Appendix.

\section{The large-scale kSZ effect} \label{sec:ksz}

\begin{figure}[htbp]
\begin{center}
	\includegraphics[width=8cm]{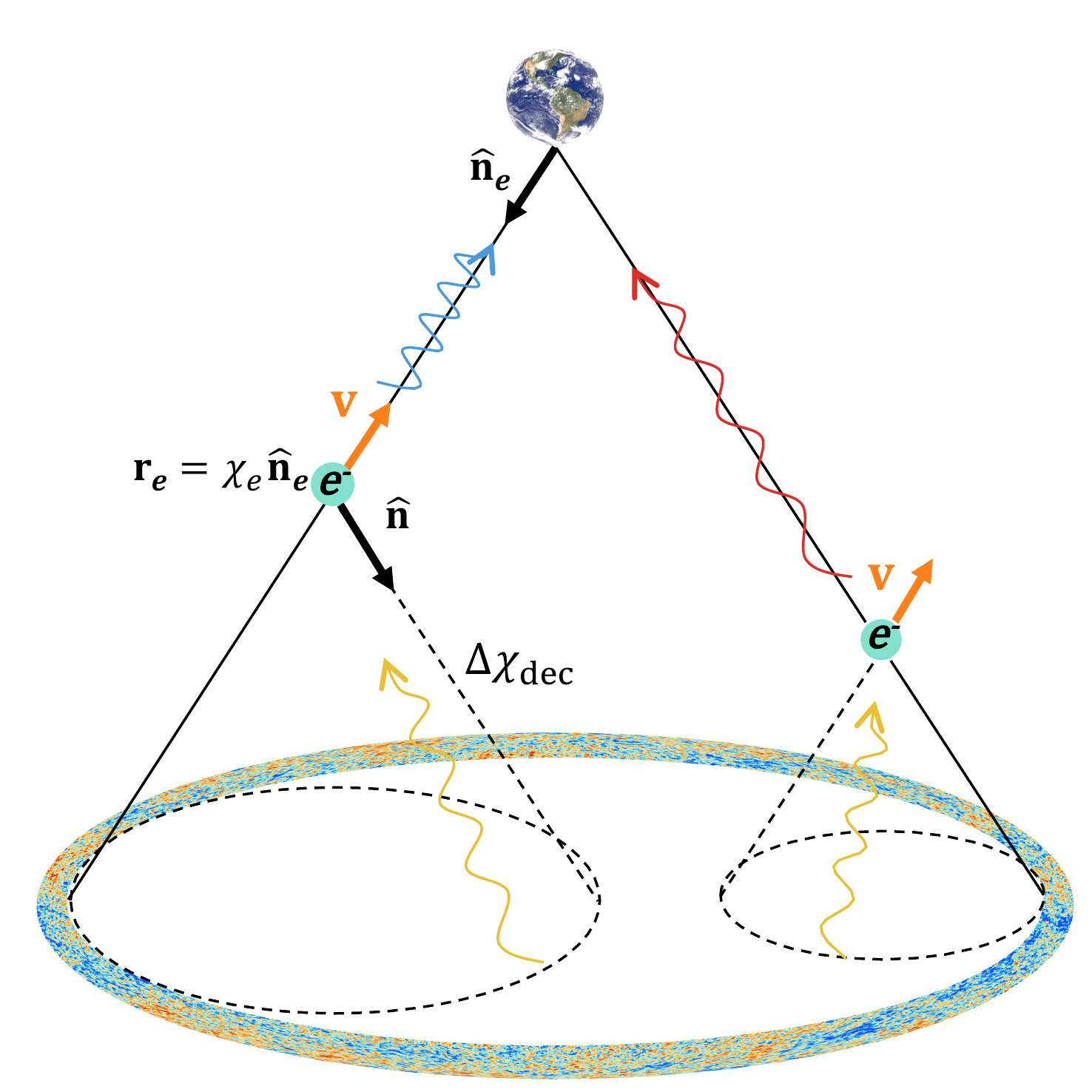}
	\caption{Scattering of CMB photons off free electrons on our past light cone. The position of an electron is described in terms of its direction ${\bf \hat{n}}_e$ and comoving distance $\chi_e$. The direction from the electron to a point on the surface of last scattering is denoted by ${\bf \hat{n} }$ and the distance to last scattering by $\Delta \chi_\text{dec}$.}
	\label{fig:lightcone}
	\end{center}
\end{figure}

The kinetic Sunyaev Zel'€™dolvich effect arises from Compton scattering of CMB photons by free electrons moving with respect to the CMB rest frame. This produces temperature anisotropies given by an integral along the line of sight:
\begin{align}\label{eq:kszlineofsight}
\frac{\Delta T}{T}\bigg|_\text{kSZ}({\bf \hat{n}}_e) & = - \sigma_T \int_0^{\chi_{\rm re}} d\chi_e \ a_e (\chi_e) \ n_e({\bf \hat{n}}_e, \chi_e) \ {\bf v}_\text{eff}({\bf \hat{n}}_e, \chi_e)  \cdot  {\bf \hat{n}}_e \\
& = - \sigma_T \int_0^{\chi_{\rm re}} d\chi_e \ a_e (\chi_e) \ \bar{n}_e(\chi_e)\ (1+\delta({\bf \hat{n}}_e, \chi_e)) \ v_{\rm eff}({\bf \hat{n}}_e, \chi_e) . \label{eqn:kSZ}
\end{align}
The geometry is depicted in figure~\ref{fig:lightcone}. In eq.~\eqref{eq:kszlineofsight},  $\sigma_T$ is the Thomson cross-section, $n_e({\bf \hat{n}}_e, \chi_e)$ is the electron number density, ${\bf \hat{n}}_e$ is the angular direction on the sky to the scatterer, and $\chi_e$ is the comoving radial coordinate to the scatterer along our past light cone,
\begin{equation}
\chi_e =  \int_0^{z_e} \frac{d z}{H(z)} =  -\int_{1}^{a_e} \ \frac{da}{H(a)a^2} ,
\end{equation}
where $z_e$ and $a_e$ are the scatterer's redshift and scale factor respectively. Below, we will use $\chi_e$ and $z_e$ interchangeably. In the second line of \eqref{eqn:kSZ}, we have written the electron number density as $n_e({\bf \hat{n}}_e, \chi_e) = \bar{n}_e(\chi_e)(1+\delta({\bf \hat{n}}_e, \chi_e))$ in terms of the average electron number density $\bar{n}_e(\chi_e)$, and the density perturbation $\delta$, and replaced ${\bf v_\text{eff}}({\bf \hat{n}}_e, \chi_e)  \cdot  {\bf \hat{n}}_e$ with the projection along the line of sight $v_{\rm eff}({\bf \hat{n}}_e, \chi_e)$. 

The quantity denoted by $v_{\rm eff}({\bf \hat{n}}_e, \chi_e)$ is the CMB dipole observed by each electron, projected along the line of sight:
\be\label{eq:veffdef}
v_{\rm eff}({\bf \hat{n}}_e, \chi_e)=\frac{3}{4\pi}\int d^2{\bf \hat{n}} \ \Theta({\bf \hat{n}}_e, \chi_e, {\bf \hat{n}}) \ ({\bf \hat{n}}\cdot{\bf \hat{n}}_e) ,
\ee
where, for a freely falling electron at position ${\bf r}_e \equiv \chi_e{\bf \hat{n}}_e$, the CMB temperature it sees along the direction ${\bf \hat{n}}$ is given by
\be \label{eq:electron_CMB}
\Theta({\bf \hat{n}}_e, \chi_e, {\bf \hat{n}}) = \Theta _{\rm SW} ({\bf \hat{n}}_e, \chi_e, {\bf \hat{n}}) + \Theta_{\rm Doppler} ({\bf \hat{n}}_e, \chi_e, {\bf \hat{n}}) + \Theta _{\rm ISW}({\bf \hat{n}}_e, \chi_e, {\bf \hat{n}}) .
\ee
The three contributions come from the Sachs-Wolfe (SW) effect generated by the gravitational potential on the LSS, the Doppler effect due to peculiar motion of photons on the LSS and peculiar motion of electrons at redshift $z_e$, and the integrated Sachs-Wolfe (ISW) effect. 

Working in Newtonian gauge
\begin{equation}
ds^2=-(1+2\Psi)dt^2+a^2(t)(1-2\Psi)d{\bf x}^2 ,
\end{equation}
the Sachs-Wolfe contribution is given by
\be \label{eq:thetaSW}
\Theta _{\rm SW} ({\bf \hat{n}}_e, \chi_e, {\bf \hat{n}}) = \left( 2D_\Psi(\chi_\text{dec}) -\frac{3}{2} \right) \Psi_i({\bf r}_\text{dec}),
\ee
where ${\bf r}_{\rm dec} \equiv \chi_e {\bf \hat{n}}_e + \Delta \chi_\text{dec} {\bf \hat{n}}$ with $\Delta\chi_\text{dec}=\Delta \chi (a_{\rm dec}) = -\int_{a_e}^{a_\text{dec}} da \left( {H(a)a^2} \right)^{-1}$ the distance along the electron's past light cone to decoupling. More generally, we will define
\be
	\Delta\chi(a)=  -\int_{a_e}^a \frac{da}{H(a)a^2} .
\ee
In eq.~\eqref{eq:thetaSW} we have used the growth function, $D_\Psi(a)$, which relates the potential to its primordial value at $a \rightarrow 0$ through the definition $\Psi({\bf r},a) =D_\Psi(a)\Psi_i({\bf r})$. The growth function is well approximated on superhorizon scales by
\be
\label{eqn:PhiSH}
D_\Psi(a)\equiv \frac{\Psi_{\rm SH}(a)}{\Psi_{\rm
  SH,i}}=\frac{16\sqrt{1+y}+9y^3+2y^2-8y-16}{10y^3} \left[ \frac{5}{2}\Omega_m \frac{E(a)}{a}\int_0^a
\frac{da}{E^3(a) \ a^3} \right] ,
\ee
where $E(a)=\sqrt{\Omega_ma^{-3}+\Omega_\Lambda}$ is the normalized Hubble parameter.

The Doppler component is given by
\be\label{eq:thetaDopp}
 \Theta_{\rm Doppler} ({\bf \hat{n}}_e, \chi_e, {\bf \hat{n}})={\bf \hat{n}}\cdot
[{\bf v}({\bf r}_e,\chi_e)-{\bf v}({\bf r}_{\rm dec},\chi_\text{dec})].
\ee
The velocities can be related to the potential through
\be
\label{eqn:vSH}
{\bf v}=-\frac{2a^2c^2 H(a)}{H^2_0\Omega_m} \frac{y}{4+3y}
\left[{\bf \nabla} \Psi+\frac{d{\bf \nabla}\Psi}{d\ln a}\right] ,
\ee
which is valid on all scales. On large scales, we can use this expression to define a velocity growth function $D_v(a)$:
\be
{\bf v}= -\frac{2a^2c^2H(a)}{H^2_0\Omega_m}  \frac{y}{4+3y}
\left[D_\Psi+\frac{dD_\Psi}{d\ln a}\right]{\bf \nabla} \Psi_{i}, 
\ee
where
\be
\label{eqn:Dv}
D_v(a)\equiv \frac{2a^2H(a)}{H^2_0\Omega_m} \frac{y}{4+3y}
\left[D_\Psi+\frac{dD_\Psi}{d\ln a}\right].
\ee

Finally, the ISW term is given by
\be
\label{eqn:ISW}
\Theta _{\rm ISW} ({\bf \hat{n}}_e, \chi_e, {\bf \hat{n}})=2\int_{a_{\rm
    dec}}^{a_e} \frac{d\Psi}{da}({\bf r}(a),a) da
=2\int_{a_{\rm
    dec}}^{a_e} \frac{dD_\Psi}{da}\Psi_i({\bf r}(a)) da .
\ee
Here, ${\bf r}(a)={\bf r}_e+\Delta \chi (a) \ {\bf \hat{n}}$.

\subsection{Fourier kernel for the effective velocity}

Relating each contribution to the effective velocity to the primordial potential $\Psi_i$ allows us to define a kernel in Fourier space relating $\Psi_i$ to the effective velocity. The details can be found in Appendix \ref{app:veff}, which results in the expression
\begin{equation}\label{eq:final_veff}
v_{\rm eff}({\bf \hat{n}}_e, \chi_e) = i \int \frac{d^3 k}{(2 \pi)^3} \ T(k) \tilde{\Psi}_i ({\bf k}) \  \mathcal{K}^v(k,\chi_e) \ \mathcal{P}_{1} ({\bf \hat{k}} \cdot  {\bf \hat{n}}_e ) \ e^{i \chi_e {\bf k} \cdot {\bf \hat{n}}_e },
\end{equation}
where we have incorporated the transfer function $T(k)$ for the potential to account for sub-horizon evolution on small scales and the kernel $\mathcal{K}^v(k,\chi_e)$ receives contributions from the SW, ISW, and Doppler terms
\begin{equation}
\mathcal{K}^v(k,\chi_e) \equiv \left[ \mathcal{K}_{\rm D} (k, \chi_e) + \mathcal{K}_{\rm SW} (k, \chi_e) + \mathcal{K}_{\rm ISW} (k, \chi_e)\right],
\end{equation}
given by
\ba\label{eq:kernel}
\mathcal{K}_{\rm D} (k, \chi_e) &\equiv& k D_v (\chi_\text{dec}) j_{0} (k \Delta \chi_\text{dec} ) - 2 k D_v (\chi_\text{dec}) j_{2} (k \Delta \chi_\text{dec} ) - k D_v (\chi_e),  \\ \label{eq:kernel2}
\mathcal{K}_{\rm SW} (k, \chi_e) &\equiv& 3\left( 2D_\Psi(\chi_\text{dec}) -\frac{3}{2} \right) j_{1} (k \Delta \chi_\text{dec}), \\ \label{eq:kernel3}
\mathcal{K}_{\rm ISW} (k, \chi_e) &\equiv& 6  \int_{a_{\rm dec}}^{a_e} da \frac{dD_\Psi}{da}  \ j_{1} (k \Delta \chi (a)) .
\ea

The SW and ISW kernels have support predominantly on large scales, while the Doppler kernel has support on all scales. The last term in the Doppler kernel, $k D_v (\chi_e)$, yields the ``conventional" kSZ effect, and represents the dominant contribution to the kSZ effect on scales that have currently been measured (e.g. using pairwise cluster velocities). Using the fact that $j_{1} (k \Delta \chi (a))$ and $j_{2} (k \Delta \chi (a))$ have support predominantly on scales $k \sim 1 / \Delta \chi$, we can estimate the order of magnitude of scales that contribute to these terms. In the range of redshift between $6>z_e>0$, we have $1.25 < H_0 \Delta \chi_\text{dec} < 3.18$. Using $k = 1 / \Delta \chi_\text{dec} $, this translates into the range $(14.3 \ {\rm Gpc} )^{-1} < k < (5.6 \ {\rm Gpc} )^{-1}$.

There is one important physical condition that must hold: a pure potential gradient should not contribute to an observable like the kSZ effect~\cite{Erickcek08}. A pure gradient  can be removed in linear perturbation theory by performing a special conformal transformation on the spatial metric. More generally, it is always possible to remove the gradient at a point by the same special conformal transformation. We include a proof of these statements in Appendix~\ref{app:gradientisgauge}. This absence of a gradient contribution to the kSZ effect has important implications for the behavior of the effective velocity Fourier kernel $\mathcal{K}^v$ at small $k$. 

To see this, consider a Newtonian potential that is a pure gradient
\begin{equation}
\Psi_i ({\bf x}) = A_j x^j .
\end{equation}
Using the properties of the derivative of the Dirac delta function, we can write this in Fourier space as
\begin{equation}
\Psi_i ({\bf k}) = i (2\pi)^3 A^j \ \partial_j \delta^3({\bf k}).
\end{equation}
Evaluating eq.~\eqref{eq:final_veff}, we obtain
\begin{equation}
v_{\rm eff}({\bf \hat{n}}_e, \chi_e) = - A^j \partial_j \mathcal{K}^v(k=0,\chi_e),
\end{equation}
where we have used the fact that $\mathcal{K}^v(k=0,\chi_e) = 0$, $T(k=0) = 1$, and $\partial_j T(k=0) = 0$. Unless $\partial_j \mathcal{K}^v(k=0,\chi_e) = 0$, there will be an observable kSZ effect from a pure gradient, which would be unphysical. This, together with the fact that each of the three contributions to $\mathcal{K}^v(k,\chi_e)$ are odd functions of $k$, implies that we must have
\begin{equation}
\mathcal{K}^v(k \rightarrow 0,\chi_e) = \mathcal{O}(k^3) + \ldots
\end{equation}

\begin{figure}[htbp]
	\begin{center}
		\subfigure{\includegraphics[width=7.5cm]{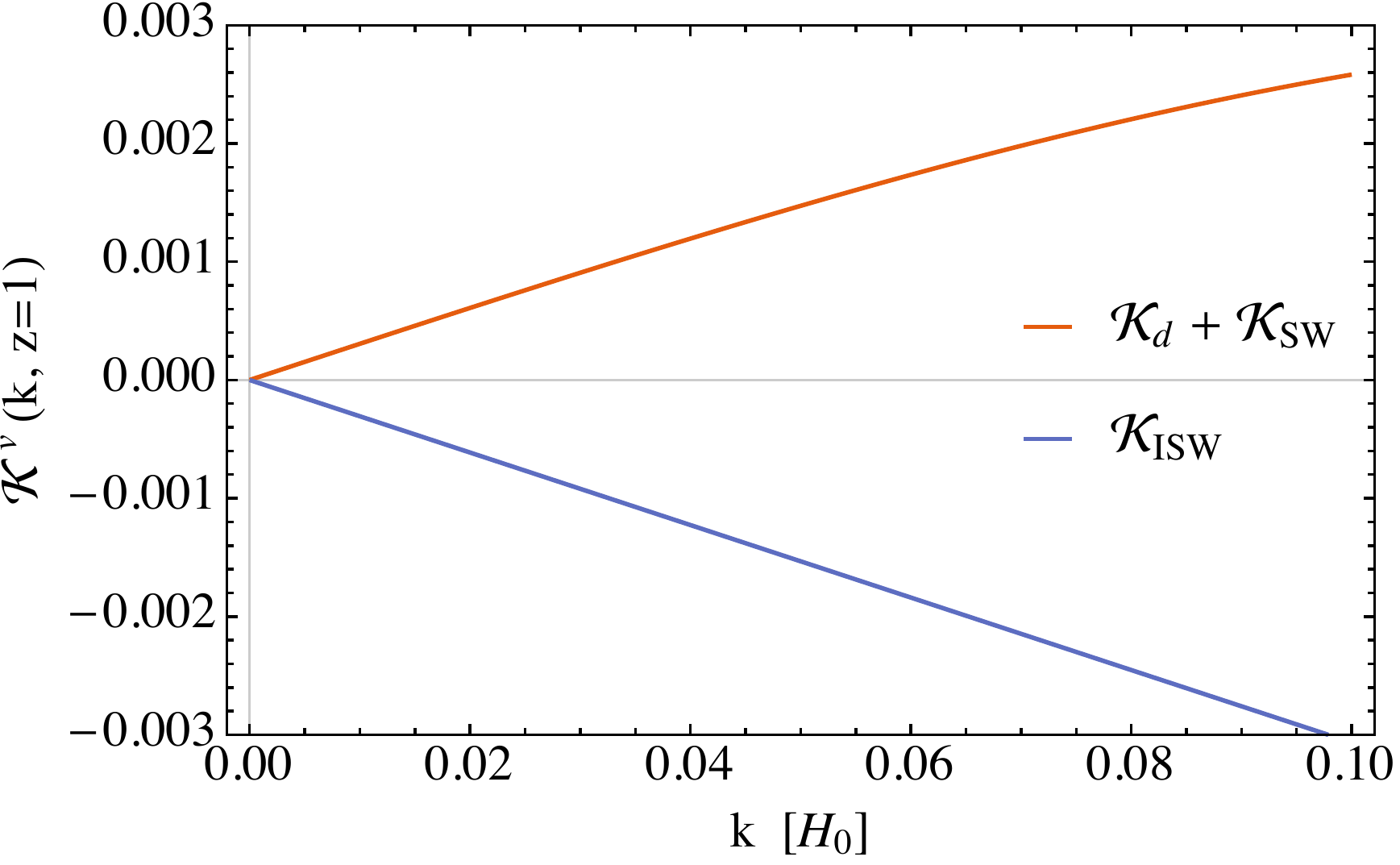}}
		\hspace{.2cm}
		\subfigure{\includegraphics[width=7.5cm]{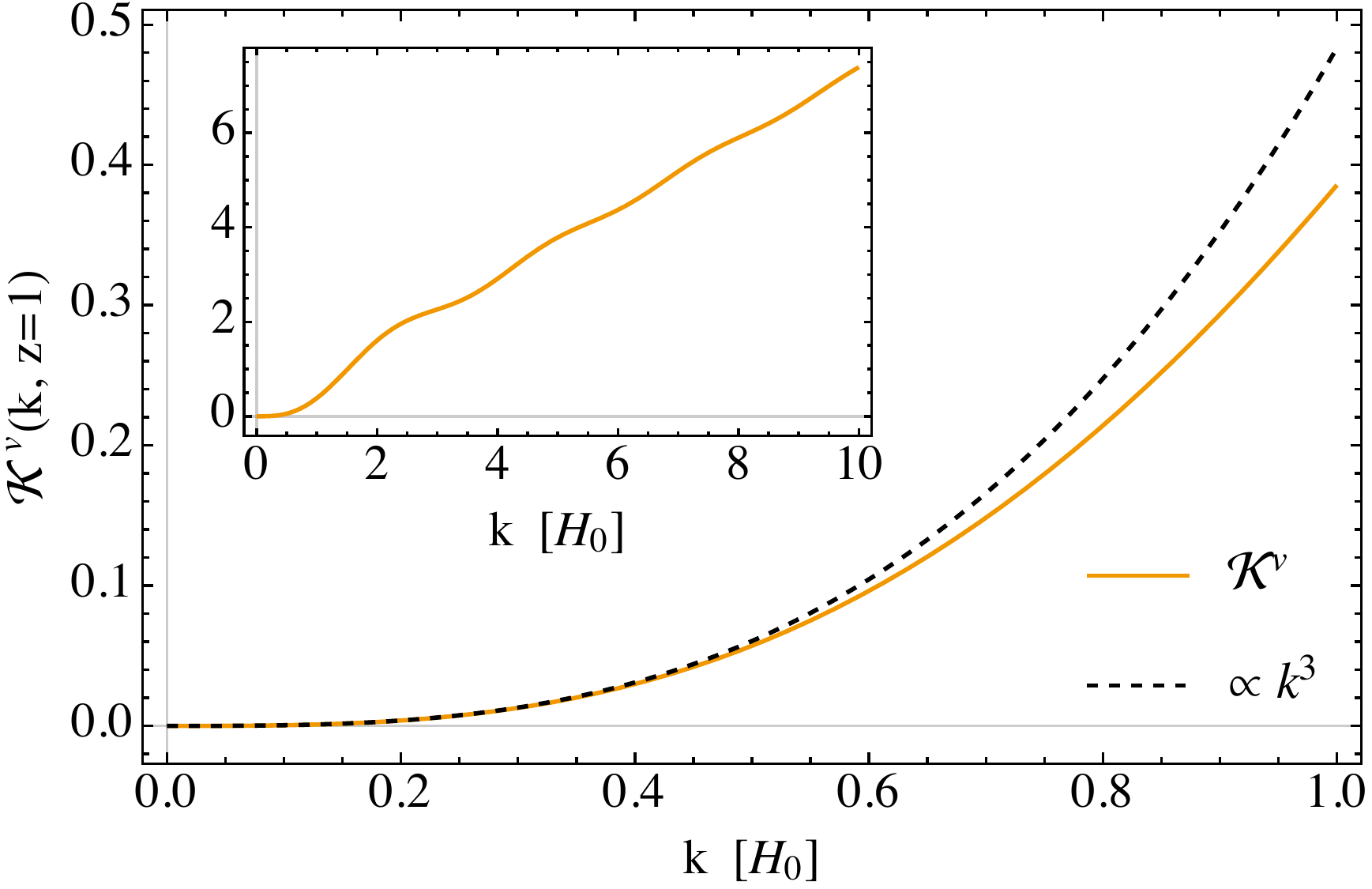}}
		\caption{The cancellation at linear order of the three pieces of $\mathcal{K}^v$ as $k \rightarrow 0$ in $\Lambda$CDM. This cancellation can be shown analytically in a universe without radiation (see Appendix \ref{sec:appB}). The leading order behavior is cubic as $k \rightarrow 0$ as shown in the right panel. }
		\label{fig:kernel}
	\end{center}
\end{figure}

Expanding $\mathcal{K}_{\rm SW}$, $\mathcal{K}_{\rm ISW}$, and $\mathcal{K}_{\rm D}$ separately, the leading order term in the Taylor series expansion is linear in $k$. Therefore, a cancellation between these terms must occur in the limit $k \rightarrow 0$. This is the same type of cancellation demonstrated for the primary CMB dipole due to a pure gradient in Ref.~\cite{Erickcek08}. We demonstrate this cancellation analytically in a Universe with matter and a cosmological constant in Appendix~\ref{sec:appB}. The cancellation, and the behavior of the full kernel $\mathcal{K}^v(k,\chi_e)$ at $z=1$ for small $k$ is shown in figure~\ref{fig:kernel}. 

Looking at the inset of the right panel in figure~\ref{fig:kernel}, we see that the contributions from the spherical Bessel functions in the SW, ISW, and Doppler kernels are evident as oscillations on the linear Doppler contribution. These effects are significant on scales up to of order $k \sim 10 \ H_0 \sim 2 \times 10^{-3} \ {\rm Mpc}^{-1}$.

\subsection{Angular decomposition of the effective velocity} \label{sec:almv}

The multipole moments of the effective velocity are given by integrating over ${\bf \hat{n}}_e$ 
\begin{align}
	a_{\ell m}^{v} (\chi_e) =& \int d^2{\bf \hat{n}}_e \ v_\text{eff}({\bf \hat{n}}_e,\chi_e)\ Y^*_{\ell m}({\bf \hat{n}}_e) \\ \label{eq:almv1}
	= &\  i \int \frac{d^3 k}{(2 \pi)^3}\ T(k)\ \tilde{\Psi}_i(\mathbf{k})\ \mathcal{K}^v(k,\chi_e) \int d^2{\bf \hat{n}}_e\ Y^*_{\ell m}({\bf \hat{n}}_e)\  \mathcal{P}_1({\bf \hat{k}}\cdot {\bf \hat{n}}_e)\ e^{i \chi_e {\bf k} \cdot {\bf \hat{n}}_e } .
\end{align}
The second integral can be written as a triple product of spherical harmonics by expanding the exponential using \eqref{eq:expidentity} and writing the Legendre polynomials in terms of spherical harmonics using
\begin{equation} \label{eq:PYY}
	\mathcal{P}_\ell ({\bf \hat{x}}\cdot{\bf \hat{x}}')=\frac{4\pi}{2\ell +1} \sum_{m=-\ell }^\ell Y^*_{\ell m}({\bf \hat{x}})\ Y_{\ell m}({\bf \hat{x}}').
\end{equation}

The integral over ${\bf \hat{n}}_e$ then becomes 
\begin{align}
	& \int  d^2{\bf \hat{n}}_eY^*_{\ell m}({\bf \hat{n}}_e) \left[ \frac{4\pi}{3} \sum_{m''=-1}^1 Y^*_{1m''}({\bf \hat{k}})Y_{1m''}({\bf \hat{n}}_e)\right]\left[ 4\pi i^{\ell'}j_{\ell'}(k\chi_e)\sum_{m'=-\ell'}^{\ell'} Y_{\ell'm'}({\bf \hat{k}})Y^*_{\ell'm'}({\bf \hat{n}}_e)\right] \no \\
	& = \frac{(4\pi)^2}{3} \sum_{\ell', m',m''} i^{\ell'} j_{\ell'}(k\chi_e)Y^*_{1m''}({\bf \hat{k}})Y_{\ell'm'}({\bf \hat{k}}) \int d^2{\bf \hat{n}}_eY^*_{\ell m}({\bf \hat{n}}_e)Y^*_{\ell'm'}({\bf \hat{n}}_e)Y_{1m''}({\bf \hat{n}}_e). \label{eq:tripleY}
\end{align}
The triple product integral can be expressed in terms of the Clebsch-Gordan coefficients $\text{C}^{\ell_1\ell_2\ell_3}_{m_1 m_2 m_3}$:
\begin{equation} \label{eq:YYY}
	 \int d^2{\bf \hat{n}}_e\ Y^*_{\ell m}({\bf \hat{n}}_e)\  Y^*_{\ell'm'}({\bf \hat{n}}_e) \ Y_{1m''}({\bf \hat{n}}_e) = \sqrt{\frac{(2\ell+1)(2\ell'+1)}{12\pi}} \ \text{C}^{\ell \ell' 1}_{0 0 0} \ \text{C}^{\ell \ell' 1}_{m m' m''}.
\end{equation}
These first coefficients $\text{C}^{\ell \ell' 1}_{0 0 0}$ are only nonzero for $\ell'=\ell\pm 1$. The second coefficients $\text{C}^{\ell\ \ell\pm 1\ 1}_{m m' m''}$ then require that $m'=m''-m$ for $m''= -1,0,1$. Therefore, the sums over $\ell'$ and $m'$ in \eqref{eq:tripleY} will select six non-zero terms in which $(\ell',m')$ take the values $(\ell+1, 1-m),\ (\ell+1, -m),\ (\ell+1, -1-m),\ (\ell-1, 1-m),\ (\ell-1, -m),\ (\ell-1, -1-m)$. Further, spherical harmonic identities and spherical Bessel recursion relations can then be used to simplify these six terms into just two terms proportional to $Y^*_{\ell m}({\bf \hat{k}})$. Equation \eqref{eq:tripleY} reduces to
\begin{equation}
	\frac{4 \pi}{2 \ell + 1} Y^*_{\ell m}({\bf \hat{k}})  \left[ i^{\ell-1}\ \ell\ j_{\ell - 1} (k \chi_e) + i^{\ell+1}(\ell + 1)  j_{\ell + 1} (k \chi_e) \right] .
\end{equation}
Plugging this in for the second integral in equation \eqref{eq:almv1} leads to the expression 
\begin{equation} \label{eq:almv}
	a_{\ell m}^{v} (\chi_e) = \int \frac{d^3 k}{(2 \pi)^3}\ \Delta_\ell^{v}(k,\chi_e) \tilde{\Psi}_i(\mathbf{k})Y^*_{\ell m}({\bf \hat{k}}),
\end{equation}
where we have defined the transfer function $\Delta^{v}_{\ell} (k, \chi_e)$ as
\be\label{eq:transfer}
\Delta^{v}_{\ell}(k, \chi_e) \equiv \frac{4\pi \ i^{\ell}}{2 \ell + 1} \mathcal{K}^v(k,z\chi_e) \left[ \ell\ j_{\ell - 1} (k \chi_e) - (\ell + 1)  j_{\ell + 1} (k \chi_e) \right] T(k).
\ee

The asymptotic behavior of the transfer function as $k \rightarrow \infty$ and $k \rightarrow 0$ is given by:
\ba
\lim_{k \rightarrow \infty} \Delta^{v}_{\ell}(k, \chi_e) &=&  -4 \pi i^\ell \frac{D_v (\chi_{e})}{\chi_e} T(k) \cos \left[k \chi_e - \ell \pi / 2\right] ,\\
\lim_{k \rightarrow 0} \Delta^{v}_{\ell}(k, \chi_e) &=& \frac{4\pi \ i^{\ell}}{2 \ell + 1}  \left[ \frac{ \ell \sqrt{\pi} (k \chi_e)^{\ell - 1}}{2^\ell \Gamma[\frac{1}{2} + \ell]} \right] c_3(\chi_e) k^3 ,
\ea
where in the small-$k$ limit we have used the fact that $T(0)=1$ and written the coefficient of the leading order (cubic) term in the Taylor series expansion of $\mathcal{K}^v(k,z\chi_e)$ as $c_3(\chi_e)$; in the large-$k$ limit, $T(k) \propto k^{-2}$.\\

\section{Simulations}\label{sec:simulations}

\begin{figure}[htbp]
  \begin{center}
			\subfigure{\includegraphics[height=6.3cm]{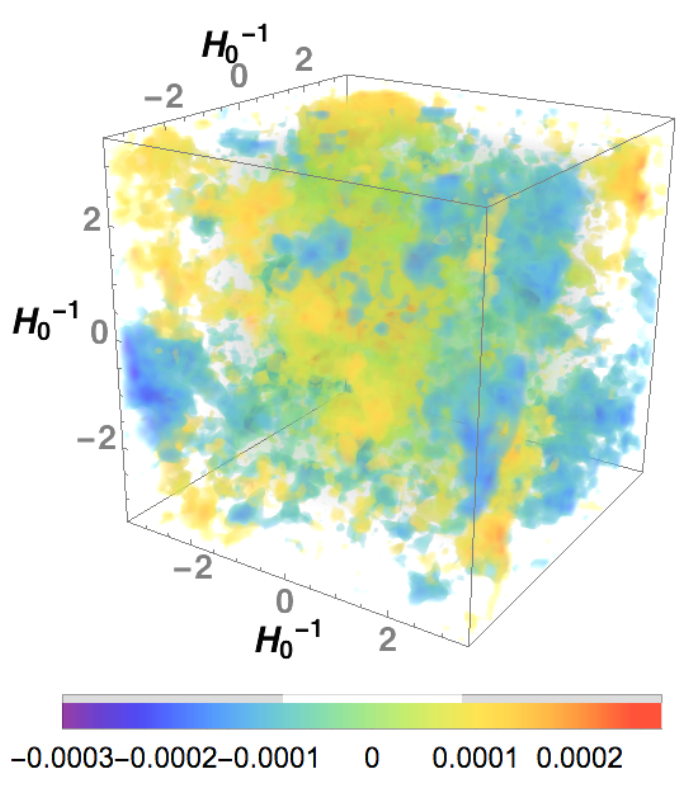}}
			\hspace{.5cm}
			\subfigure{\includegraphics[height=6.3cm]{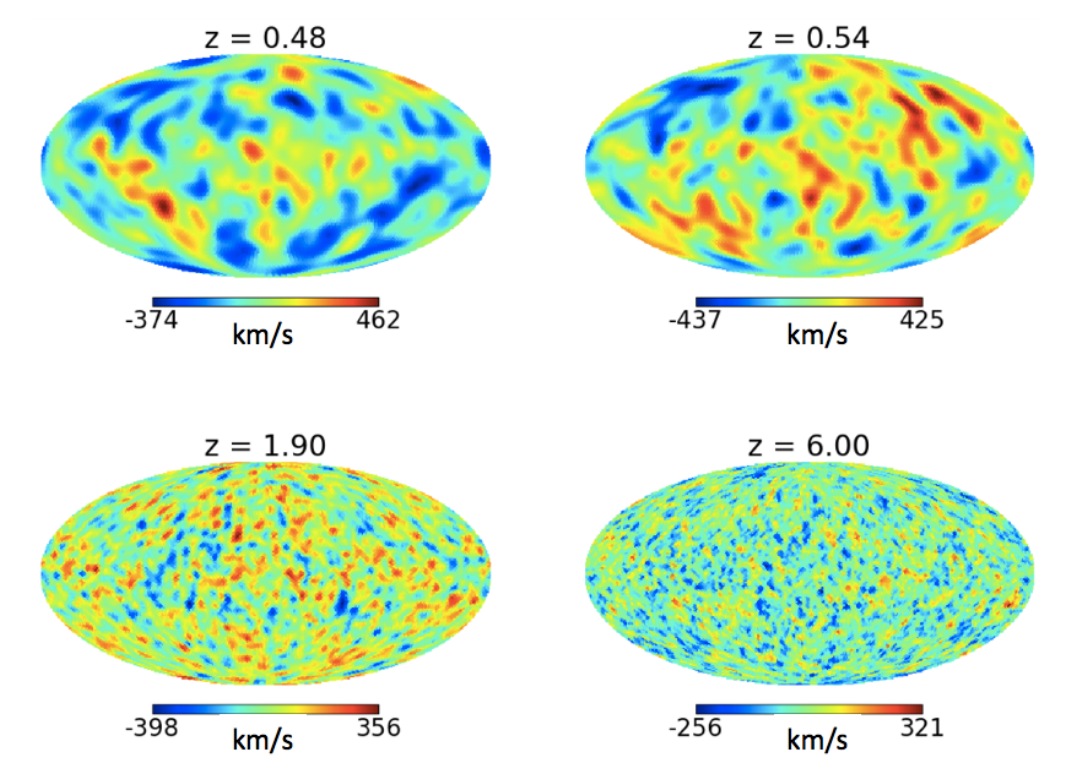}}
			\caption{By generating random realizations of $\tilde{\Psi}_i({\bf k})$, drawn from a Gaussian distribution with power $P_\Psi(k)$, we can construct realizations for $v_\text{eff}$. The left panel shows an example realization for $\Psi_i({\bf r})$, and the resulting maps of $v_\text{eff}$ at various redshifts, using a resolution of $k_\text{max}\sim57\ H_0$ ($\lambda_\text{min}\sim 484$ Mpc). Notice that correlations are evident between the top two $v_\text{eff}$ maps at close redshift, but not apparent for widely separated redshifts. Also note the increase in structure at higher redshift.}
			 \label{fig:real}
		\end{center} 
\end{figure}

In this section we describe a suite of simulations used to explore the large-scale kSZ effect and provide a concrete example of the relation between the primordial gravitational potential and the effective velocity. This will be used to compute the kSZ signal-to-noise in section~\ref{sec:realizations}. We create three dimensional realizations of the primordial gravitational potential $\Psi({\bf x})$ consistent with $\Lambda$CDM using the method described in Ref.~\cite{0067-0049-137-1-1} and reviewed in Appendix~\ref{sec:randomfields}. The box size used in each case was $L = 7 H_0^{-1} \simeq 31.3 \ {\rm Gpc}$. One hundred realizations were created at a resolution of $128^3$, covering scales down to $k_{\rm max} \simeq 57.4 \ H_0$ ($\lambda_{\rm min} \simeq 484 \ {\rm Mpc}$). An example realization is shown in figure~\ref{fig:real}.

With a set of realizations in hand, we then place a hypothetical observer at the center of the box and generate $v_{\rm eff}({\bf \hat{n}}_e, \chi_e)$ at $50$ equally spaced values of $\chi_e$ at a Healpix resolution~\cite{2005ApJ...622..759G} of $N_{\rm side} = 32$ ($12,288$ equal area pixels of approximately $3.36$ square degrees each). This is done as follows. First, we write the effective velocity as
\begin{equation}\label{eq:veffcomponents}
v_{\rm eff}({\bf \hat{n}}_e, \chi_e) = i {\bf \hat{n}}_e \cdot {\bf V}({\bf \hat{n}}_e,\chi_e) ,
\end{equation}
where 
\begin{equation}
{\bf V}({\bf \hat{n}}_e,\chi_e) \equiv \int \frac{d^3 k}{(2 \pi)^3} \ \left[ T(k) \tilde{\Psi}_i ({\bf k}) \  \mathcal{K}^v(k,\chi_e) \ \frac{\bf k}{k} \right] \ e^{i \chi_e {\bf k} \cdot {\bf \hat{n}}_e } .
\end{equation}
The three components of ${\bf V}({\bf \hat{n}}_e,\chi_e)$ can be straightforwardly evaluated for each realization, at each redshift, using an fast Fourier transform (FFT) algorithm. Plugging back into eq.~\eqref{eq:veffcomponents} and choosing ${\bf \hat{n}}_e \cdot {\bf \hat{z}} = \cos \theta$, ${\bf \hat{n}}_e \cdot {\bf \hat{x}} = \sin \theta \cos \phi$, and ${\bf \hat{n}}_e \cdot {\bf \hat{y}} = \sin \theta \sin \phi$, we interpolate the resulting $v_{\rm eff}({\bf \hat{n}}_e, \chi_e)$ at each $\chi_e$ onto the Healpix grid. We then take advantage of the Healpix fast spherical harmonic transform functionality to obtain $a_{\ell m}^{v} (\chi_e)$ at each redshift in each realization.
  
In the right panel of figure~\ref{fig:real}, we show $v_{\rm eff}({\bf \hat{n}}_e, \chi_e)$ at a variety of redshifts in a single realization. In the top row, we choose two nearby redshifts, where it can be seen by eye that there is a good deal of correlation between the two maps. This is because the same large-scale potential field is responsible for the effective velocities at nearby redshifts. In the bottom row, we choose fairly distant redshifts, where the correlation between the two maps is largely absent. Note also the increasing structure with redshift. This is partially due to the limited resolution of the simulation in this figure, but more physically, there is a real effect due to the redshift transfer function. Based on the smallest structures in the resolution probed in our simulation, $64$ radial samples would capture all radial structures. Empirically, for all but the smallest angular scale structures, the coherence length between redshift slices is sufficiently long to justify our choice of $50$ values of $\chi_e$.

\section{kSZ Tomography} \label{sec:signal}

There is a large amount of information lost in performing the line-of-sight integral in eq.~\eqref{eq:kszlineofsight} for the global kSZ signal. One can in principle do far better by cross correlating the kSZ temperature anisotropies with tracers of the electron density field of known redshift. This is evident in the first detections of the kSZ effect, which were made by isolating the component of the temperature anisotropies associated with the pairwise motion of clusters, whose hot interiors harbor a large density of free electrons. In what follows, we assume the most optimistic scenario possible, in which we have perfect knowledge of the electron density field obtained through the measurement of a completely unbiased tracer. We further assume a purely Gaussian primordial power spectrum, consistent with the current constraints from Planck~\cite{Planck:2015}. 

\begin{figure}[htbp]
	\begin{center}
	\includegraphics[width=8cm]{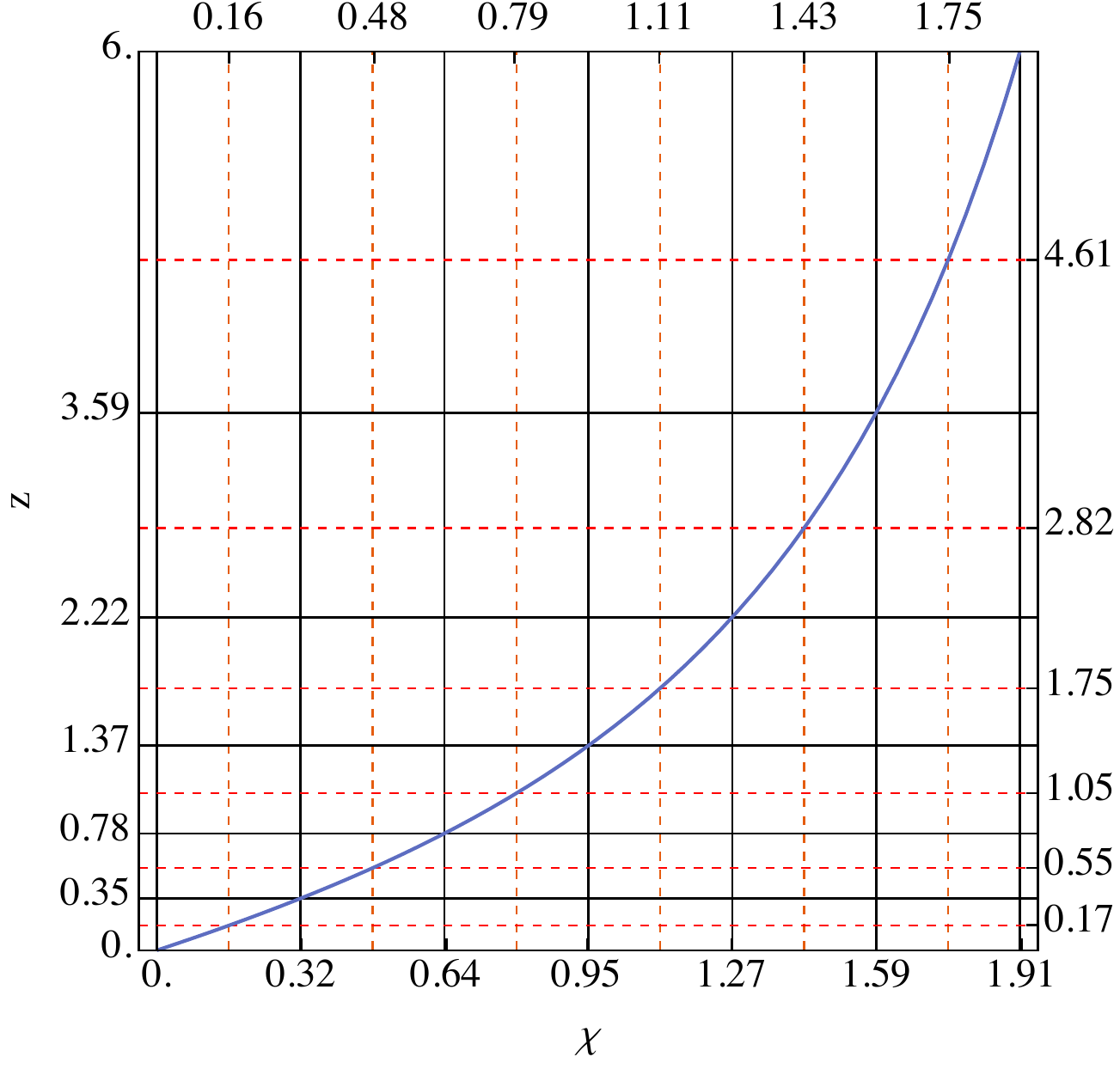}
	\caption{Redshift is displayed as a function of comoving distance to illustrate the redshift bin configurations. The solid grid lines show the boundaries that define the redshift bins for $N_\text{bin}=6$. The 12 bin configuration is represented by including the dashed grid lines. It is easy to infer the boundary values for $N_\text{bin}=24$.}
	\label{fig:bins}
	\end{center}
\end{figure}

To tease out the redshift dependence of the large scale kSZ effect, we introduce a window function $W(\chi_e, \bar{\chi}_e)$ that gives the electron density in a set of redshift bins centered on $\chi_e = \bar{\chi}_e$
\begin{equation}
\delta({\bf \hat{n}}_e, \bar{\chi}_e)=\int d\chi_e W(\chi_e, \bar{\chi}_e)\delta({\bf \hat{n}}_e, \chi_e).
\end{equation}
In the following, we use a top-hat window function normalized to unity: $\int_0^{\chi_\infty} d\chi W(\chi, \bar{\chi})=1$. We will consider scenarios with $6$, $12$, and $24$ redshift bins of equal width, covering the range $0 < z < 6$. The redshift coverage for each bin configuration is shown in Figure~\ref{fig:bins}. 

Forming the cross correlation between the kSZ contribution to the CMB temperature anisotropies and the windowed electron density field, we obtain
\begin{align}
	\bang{ \frac{\Delta T}{T}\bigg|_\text{kSZ}({\bf \hat{n}}_e)\ \delta({\bf \hat{n}}'_e,  \bar{\chi}_e)} = & \sigma_T \int d\chi'_e\ W(\chi'_e, \bar{\chi}_e)\ \int d\chi_e \  a(\chi_e) \ \bar{n}_e(\chi_e) \no \\
	& \times \bang{(1+ \delta({\bf \hat{n}}_e, \chi_e)) v_{\rm eff}({\bf \hat{n}}_e, \chi_e)  \delta({\bf \hat{n}}_e', \chi'_e)} .
	\label{eq:crosspower}
\end{align}

Now comes a very important step. The correlation function above is defined as an ensemble average. Typically, one is interested in using the measured correlation functions to constrain a statistical model of the ensemble. Here, this is not the case. Instead, {\em we strive to learn information about our particular realization} using the cross correlation, which is information that in the former scenario would have been an obstruction to learning about the theoretical model of the ensemble (e.g. cosmic variance). As we wish to learn about large-scale inhomogeneities, the ensemble average in eq.\eqref{eq:crosspower} should only be taken over small scales, leaving large scales as a fixed deterministic field.

To formulate this idea more precisely, we decompose the primordial potential into long and short wavelength fields, which can be defined in Fourier space as: 
\begin{equation}
\Psi_i^\mathcal{L} ({\bf x}) = \int \frac{d^3 k}{(2 \pi)^3} \mathcal{L} (k) \Psi_i ({ \bf k}) e^{i {\bf k} \cdot {\bf x}}, \ \  \Psi_i^\mathcal{S} ({\bf x}) = \int \frac{d^3 k}{(2 \pi)^3} \mathcal{S} (k) \Psi_i ({ \bf k}) e^{i {\bf k} \cdot {\bf x}} ,
\end{equation}
such that $\mathcal{L} (k) + \mathcal{S} (k) = 1$. A suitable choice could be $\mathcal{L} (k) = e^{- k^2 / 2 k_{*}^2 }, \ \mathcal{S} (k) = 1 - e^{- k^2 / 2 k_{*}^2 }$, although we will implicitly be choosing a step function for $\mathcal{L} (k)$ in what follows. Below, we imagine that scales larger than $k_* \agt 10^{-2} \ {\rm Mpc}^{-1}$ form the deterministic long field while smaller scales form the stochastic short field. The precise choice of $k_*$ does not affect our results because as we show in section~\ref{sec:noise} and~\ref{sec:estsig} small scales ($k \gg 10^{-2} \ {\rm Mpc}^{-1}$) do not contribute to the signal and large scales ($k \ll 10^{-2} \ {\rm Mpc}^{-1}$) do not contribute to the noise. 

Because we are working in the linear regime, a long-short split in $\Psi_i$ translates into a long-short split in the effective velocity $v_{\rm eff}$ and the electron density field $\delta$, which we therefore decompose as:
\begin{equation}
v_\text{eff} ({\bf \hat{n}}_e,\chi_e) = v_\text{eff}^\mathcal{L} ({\bf \hat{n}}_e,\chi_e) +  v_\text{eff}^\mathcal{S} ({\bf \hat{n}}_e,\chi_e), \ \ \delta({\bf \hat{n}}_e,\chi_e) = \delta^\mathcal{L} ({\bf \hat{n}}_e,\chi_e) + \delta^\mathcal{S} ({\bf \hat{n}}_e,\chi_e).
\end{equation}
Substituting these expansions into the ensemble average in eq.~\eqref{eq:crosspower} and extracting the long wavelength fields in the sense described above, we obtain:
\begin{align}
	\ang{(1+\delta)\ v_\text{eff}\ \delta'} = \ & \ang{(1 + \delta^\mathcal{L} + \delta^\mathcal{S})\ (v_\text{eff}^\mathcal{L} + v_\text{eff}^\mathcal{S})\ ({\delta'}^\mathcal{L} +{\delta'}^\mathcal{S})} \no \\
	= \ & v_\text{eff}^\mathcal{L}{\delta'}^\mathcal{L}
	+ v_\text{eff}^\mathcal{L}\delta^\mathcal{L}{\delta'}^\mathcal{L}  \nonumber \\
	& + v_\text{eff}^\mathcal{L}\ang{{\delta'}^\mathcal{S}}
	+ {\delta'}^\mathcal{L}\ang{v_\text{eff}^\mathcal{S}}
	+ \delta^\mathcal{L}{\delta'}^\mathcal{L}\ang{v_\text{eff}^\mathcal{S}}
	+ v_\text{eff}^\mathcal{L}\delta^\mathcal{L}\ang{{\delta'}^\mathcal{S}}
	+ v_\text{eff}^\mathcal{L}{\delta'}^\mathcal{L}\ang{{\delta}^\mathcal{S}}  \nonumber \\
	& + \ang{v_\text{eff}^\mathcal{S}\delta^\mathcal{S}{\delta'}^\mathcal{S}}  \nonumber \\
	& + v_\text{eff}^\mathcal{L} \ang{\delta^\mathcal{S}{\delta'}^\mathcal{S}} 
	+ \delta^\mathcal{L}\ang{v_\text{eff}^\mathcal{S}{\delta'}^\mathcal{S}}
	+ {\delta'}^\mathcal{L}\ang{v_\text{eff}^\mathcal{S}{\delta}^\mathcal{S}}	
	+ \ang{v_\text{eff}^\mathcal{S}{\delta'}^\mathcal{S}} ,
\end{align} 
where $\delta'$ represents $\delta({\bf \hat{n}}_e', \chi'_e)$. From our assumption that the short wavelength components are approximately Gaussian, we set the one-point and three-point correlation functions of short wavelength fields to zero, resulting in the final expression:\footnote{In any case, significant non-Gaussianity on small scales will not directly mimic the signal we are ultimately after, which is a {\em long wavelength} modulation of short wavelength power.} 
\begin{align}
	\ang{(1+\delta)\ v_\text{eff}\ \delta'} = \ & v_\text{eff}^\mathcal{L}{\delta'}^\mathcal{L}
	+ v_\text{eff}^\mathcal{L}\delta^\mathcal{L}{\delta'}^\mathcal{L} \no \\
	& + \ang{v_\text{eff}^\mathcal{S}{\delta'}^\mathcal{S}} \nonumber \\
	& + v_\text{eff}^\mathcal{L} \ang{\delta^\mathcal{S}{\delta'}^\mathcal{S}} 
	+ \delta^\mathcal{L}\ang{v_\text{eff}^\mathcal{S}{\delta'}^\mathcal{S}}
	+ {\delta'}^\mathcal{L}\ang{v_\text{eff}^\mathcal{S}{\delta}^\mathcal{S}} .
\end{align} 
The terms on the first line give rise to fluctuations on large angular scales, where the primary CMB dominates. We can therefore eliminate this hopelessly unmeasurable term by filtering the CMB on large angular scales $\ell \alt 3000$.\footnote{The kSZ effect dominates the CMB temperature anisotropies on scales $\ell \gtrsim 3000$, as shown in figure~\ref{fig:ClTTfull}.} The term on the second line gives rise to a statistically isotropic cross power. The terms on the third line give rise to a long wavelength modulation of small-scale power, and will be the focus of what follows. The first of these sources of power asymmetry, $v_\text{eff}^\mathcal{L} \ang{\delta^\mathcal{S}{\delta'}^\mathcal{S}}$, is far larger than the other two. If we consider a single long ($k^\mathcal{L}$) and a single short ($k^\mathcal{S}$) wavelength mode, then noting that the Doppler term dominates $v_\text{eff}^\mathcal{S}$ and using $v \propto \delta/k$, we have $\delta^\mathcal{L}\ang{v_\text{eff}^\mathcal{S}{\delta'}^\mathcal{S}} \sim (k^\mathcal{L}/k^\mathcal{S})v_\text{eff}^\mathcal{L} \ang{\delta^\mathcal{S}{\delta'}^\mathcal{S}} \ll v_\text{eff}^\mathcal{L} \ang{\delta^\mathcal{S}{\delta'}^\mathcal{S}}$. We can therefore approximate
\begin{align}
	\ang{(1+\delta)\ v_\text{eff}\ \delta'} \simeq \ & \ang{v_\text{eff}^\mathcal{S}{\delta'}^\mathcal{S}} + v_\text{eff}^\mathcal{L} \ang{\delta^\mathcal{S}{\delta'}^\mathcal{S}} , \label{eq:dvd}
\end{align} 
illustrating that there is a statistically isotropic contribution in the first term that depends only on small scales, and an anisotropic contribution in the second term that depends on large scales. The desired signal is captured in the anisotropic power asymmetry, whereas the small scale isotropic component contributes to the noise which is computed next in section~\ref{sec:noise}. Focusing here on the signal, substituting eq.~\eqref{eq:dvd} into eq.~\eqref{eq:crosspower}, and suppressing the $\mathcal{S}$ and $\mathcal{L}$ superscripts, we obtain
\begin{align}\label{eq:crosscorrelation_intermediate}
	\bang{ \frac{\Delta T}{T}\bigg|_\text{kSZ}({\bf \hat{n}}_e) \delta({\bf \hat{n}}'_e, \bar{\chi}_e)} = & \sigma_T \int d\chi_e \  a(\chi_e) \ \bar{n}_e(\chi_e) \ v_{\rm eff}({\bf \hat{n}}_e, \chi_e) \int d\chi'_e\ W(\chi'_e, \bar{\chi}_e) \no \\ 
	& \times \ang{ \delta({\bf \hat{n}}_e, \chi_e))\  \delta({\bf \hat{n}}'_e, \chi'_e)} + {\rm isotropic}.
\end{align}
Assuming the electron distribution traces the dark matter, the electron density correlation function is given by
\begin{align}
	\ang{ \delta({\bf \hat{n}}_e, \chi_e))\  \delta({\bf \hat{n}}_e', \chi'_e)} & = \sum_\ell \frac{2\ell+1}{4\pi}  C_\ell^{\delta\delta}(\chi_e,\chi'_e)  \mathcal{P}_\ell({\bf \hat{n}}_e \cdot {\bf \hat{n}}_e') ,\\
	\text{with} \ \ \ C_\ell^{\delta\delta}(\chi_e,\chi'_e) & = \int \frac{dk\ k^2}{(2\pi)^3} \ 4\pi\ j_\ell(k\chi_e) \sqrt{P_\delta(k,\chi_e)}\ 4\pi\ j_\ell(k\chi'_e) \sqrt{P_\delta(k,\chi'_e)}, \label{eq:Cldd}
\end{align}
where $P_\delta(k,\chi_e)$ is the non-linear matter power spectrum~\cite{Smith:2002dz}, which was computed using the Cosmicpy package.

We now arrive at the main conclusion of this section. By isolating the statistically anisotropic term in eq.~\eqref{eq:crosscorrelation_intermediate}, it is possible to measure $v_\text{eff}$. Because $v_\text{eff}$ is related to the primordial potential $\Psi_i$ in linear theory by eq.~\eqref{eq:final_veff}, this potentially opens a new observational window on large scale inhomogeneities in our Universe. Given a specific model for $\Psi_i$, it is possible to design an optimal filter to extract the power asymmetry described above~\cite{Zhang10d,Li12,Zhang:2015uta}. However, in the present context of random Gaussian fields, we quantify the power asymmetry by decomposing into power multipoles~\cite{Pullen:2007tu} that capture the anisotropic term in \eqref{eq:dvd},
\begin{equation} \label{eq:sigdef}
	b_{LM} (\bar{\chi}_e) =\int d^2{\bf \hat{n}}_e\ Y^*_{LM}({\bf \hat{n}}_e) \ \bang{ \frac{\Delta T}{T}\bigg|_\text{kSZ}({\bf \hat{n}}_e)\ \delta({\bf \hat{n}}_e, \bar{\chi}_e)} .
\end{equation}
Let us note that \eqref{eq:sigdef} is a complete characterization of the signal, as there is no extra information contained in the correlation with ${\bf \hat{n}}_e \neq {\bf \hat{n}}'_e$. This is clear from \eqref{eq:crosscorrelation_intermediate} in the limit of gaussian fields. The next step is to expand the long-wavelength effective velocity into multipoles (see section~\ref{sec:almv}) using $v_{\rm eff}({\bf \hat{n}}_e, \chi_e) = \sum_{\ell,m}a_{\ell m}^v(\chi_e)Y_{\ell m}({\bf \hat{n}}_e)$ where $a_{\ell m}^v$ is given by \eqref{eq:almv}. This makes the angular integral in \eqref{eq:sigdef} easy to compute as $\int d^2{\bf \hat{n}}_e Y^*_{LM}({\bf \hat{n}}_e)Y_{\ell m}({\bf \hat{n}}_e) = \delta_{\ell L}\delta_{mM}$, resulting in the expression,
\begin{align} 
	b_{LM} (\bar{\chi}_e) = & \sum_{\ell=\ell_\text{min}}^{\ell_\text{max}} \frac{2\ell+1}{4\pi} \int dk \frac{2k^2}{\pi} \int d\chi_e \  \sigma_T \ a(\chi_e) \ \bar{n}_e(\chi_e) \ a_{LM}^v(\chi_e) \sqrt{P_\delta(k,\chi_e)}\ j_\ell(k\chi_e) \no \\
	& \times  \int d\chi'_e\ W(\chi'_e, \bar{\chi}_e) \sqrt{P_\delta(k,\chi'_e)}\ j_\ell(k\chi'_e).
	\label{eq:sigreal}
\end{align}
We therefore see that the power multipoles in each bin are proportional to a weighted integral of the corresponding multipole of the projected effective velocity field. The lower and upper bounds on the summation, ($\ell_\text{min},\ \ell_\text{max}$), reflect the filtering and resolution scales that might be achievable. By default, we will use  ($\ell_\text{min}=3000,\ \ell_\text{max}=\infty$) unless otherwise stated. The effects of varying these bounds will be discussed in section~\ref{sec:detectability}.

\section{Cosmic Variance Limited Noise} \label{sec:noise}

In this section we estimate the cosmic variance limited noise that we expect for the power multipoles computed in the previous section. However, above, our focus was the statistically anisotropic contribution to the cross correlation, but here we are interested in the statistically isotropic contribution to temperature anisotropies, which depends predominantly on small scales. 

\begin{figure}[htbp]
	\begin{center}
	\includegraphics[width=9cm]{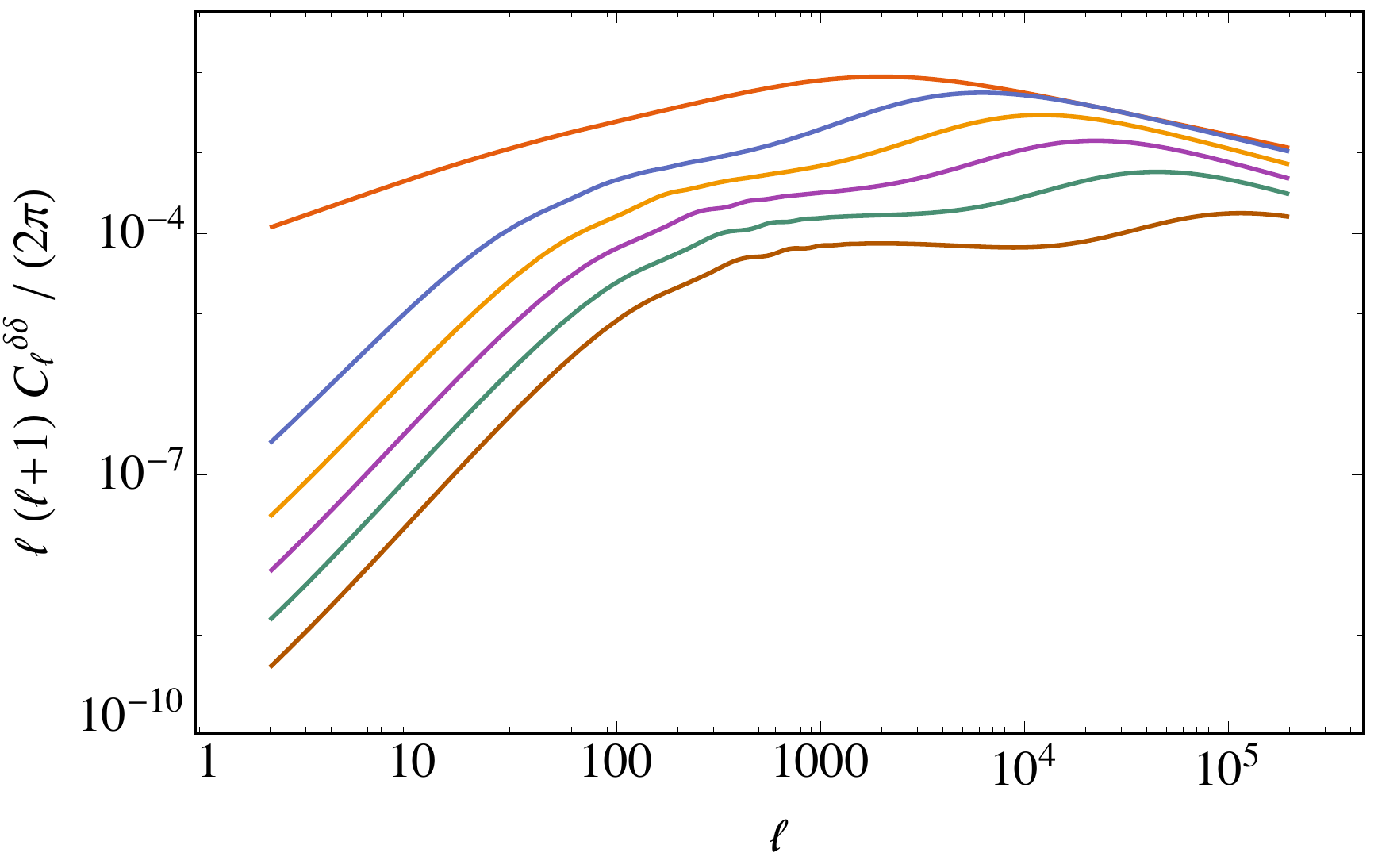}
	\caption{The non-linear angular matter power spectrum computed in six different redshift bins described by the solid lines in figure~\ref{fig:bins} from today (top) to $z=6$ (bottom).}
	\label{fig:Cldd}
	\end{center}
\end{figure}

On small angular scales, the late-time kSZ effect is the dominant source of temperature anisotropies in the CMB (see the right panel of figure~\ref{fig:ClTTfull} below). Under the assumption of approximate Gaussianity, and assuming a perfectly unbiased measurement of the electron density field, the primary source of noise is therefore an ``accidental" power asymmetry in the cross correlation. We can estimate this through the variance in the power multipoles, which is computed as the coincident limit of the four point function between two powers of $\frac{\Delta T}{T}$ and two powers of $\delta$. Specifically, we must compute:
 \begin{align}
	\ang{\tilde{b}^*_{LM} (\bar{\chi}_e) \tilde{b}_{LM} (\bar{\chi}_e)} = & \int d^2{\bf \hat{n}}_e d^2{\bf \hat{n}}'_e\  Y^*_{LM}({\bf \hat{n}}_e) \ Y_{LM}({\bf \hat{n}}'_e) \no \\
	& \times \bang{ \frac{\Delta T}{T}({\bf \hat{n}}_e)\ \delta({\bf \hat{n}}_e, \bar{\chi}_e)\ \frac{\Delta T}{T}({\bf \hat{n}}'_e)\ \delta({\bf \hat{n}}'_e, \bar{\chi}_e)} ,
	  \label{eq:noisedef}
\end{align}
where we are using tildes on the $b_{LM}$'s to indicate that these are not simply the same power multipoles as in \eqref{eq:sigreal}. Instead, the variance here captures the chance power asymmetry that is present in the statistically isotropic contribution to $\frac{\Delta T}{T}$, which is sensitive only to small scales where the ``conventional" kSZ effect dominates. Since the primary CMB also causes statistical fluctuations in the anisotropic kSZ measurement, we consider both kSZ and primary CMB contributions to $\frac{\Delta T}{T}$:
\begin{equation}
	\frac{\Delta T}{T} = \frac{\Delta T}{T}\bigg|_\text{kSZ}+\frac{\Delta T}{T}\bigg|_\text{p}
\end{equation}
The 4-point function has contributions from all possible 2-point functions and an irreducible/connected piece (which we assume to be negligibly small). Noting that the cross correlation of the primary CMB and the density field is zero, we obtain,
\begin{align}
	& \bang{ \frac{\Delta T}{T}({\bf \hat{n}}_e)\ \delta({\bf \hat{n}}_e, \bar{\chi}_e)\ \frac{\Delta T}{T}({\bf \hat{n}}'_e)\ \delta({\bf \hat{n}}'_e, \bar{\chi}_e)} \no \\
	= & \bang{ \frac{\Delta T}{T}({\bf \hat{n}}_e)\ \frac{\Delta T}{T}({\bf \hat{n}}'_e)}\bang{ \delta({\bf \hat{n}}_e, \bar{\chi}_e)\ \delta({\bf \hat{n}}'_e, \bar{\chi}_e)} \no \\
	& + \bang{ \frac{\Delta T}{T}\bigg|_\text{kSZ}({\bf \hat{n}}_e)\ \delta({\bf \hat{n}}'_e, \bar{\chi}_e)} \bang{\frac{\Delta T}{T}\bigg|_\text{kSZ}({\bf \hat{n}}'_e)\ \delta({\bf \hat{n}}_e, \bar{\chi}_e)}. \label{eq:4point}
\end{align}
Assuming the electron density field traces dark matter, the electron density autocorrelation function is given by
\begin{align}
	\bang{ \delta({\bf \hat{n}}_e, \bar{\chi}_e)\ \delta({\bf \hat{n}}'_e, \bar{\chi}_e)} = & \sum_\ell \frac{2\ell+1}{4\pi} \ C_\ell^{\delta\delta} \ \mathcal{P}_\ell({\bf \hat{n}}_e \cdot {\bf \hat{n}}'_e) ,\\
	\text{with} \ \ \ C_\ell^{\delta\delta} = & \int d\chi_e \ W(\chi_e, \bar{\chi}_e) \int d\chi'_e \ W(\chi'_e, \bar{\chi}_e) \ C_\ell^{\delta\delta}(\chi_e,\chi'_e) \no \\
	= & \int dk \frac{2k^2}{\pi} \int d\chi_e \sqrt{P_\delta(k,\chi_e)}W(\chi_e, \bar{\chi}_e)j_\ell(k\chi_e) \no \\
	& \times \int d\chi'_e \sqrt{P_\delta(k,\chi'_e)}W(\chi'_e, \bar{\chi}_e)j_\ell(k\chi'_e) \no \\
	\simeq & \int\frac{dk}{\ell+1/2} W^2\left(\frac{\ell+1/2}{k}, \bar{\chi}_e\right)P_\delta\left(k,\frac{\ell+1/2}{k}\right),
\end{align}
where we used the expression for $C_\ell^{\delta\delta}(\chi_e,\chi'_e)$ from \eqref{eq:Cldd}, and the Limber approximation \cite{Loverde:2008aa} in the last line. The quantity $C_\ell^{\delta\delta}$ is shown in figure~\ref{fig:Cldd}.

The temperature autocorrelation function has contributions from the primary CMB and the conventional kSZ effect. The latter dominates past $\ell \sim 3000$ and has the following form:
\begin{align}
	\bang{ \frac{\Delta T}{T}\bigg|_\text{kSZ}({\bf \hat{n}}_e)\ \frac{\Delta T}{T}\bigg|_\text{kSZ}({\bf \hat{n}}'_e)} = & \int d\chi_e \sigma_T a(\chi_e) \bar{n}_e(\chi_e)  \int d\chi'_e \sigma_T a(\chi'_e) \bar{n}_e(\chi'_e) \no \\  
	& \times \ang{q(\chi_e,{\bf \hat{n}}_e)q(\chi'_e,{\bf \hat{n}}'_e)}\\ 
	= &  \sum_\ell \frac{2\ell+1}{4\pi} \ C_\ell^{TT, \text{kSZ}} \  \mathcal{P}_\ell({\bf \hat{n}}_e \cdot {\bf \hat{n}}'_e) ,
\end{align}
where $q \equiv {\bf q} \cdot {\bf\hat{n}}_e \equiv v_\text{eff}(1+\delta)$ is the momentum field of free electrons. To obtain an expression for the temperature power spectrum, $C_\ell^{TT, \text{kSZ}}$, the key quantity to compute is $ \ang{q(\chi_e,{\bf \hat{n}}_e)q(\chi'_e,{\bf \hat{n}}'_e)}$, which is the fourth moment of two $\delta$'s and two $v$'s. Schematically, $\ang{qq}=\ang{vv}\ang{\delta\delta}+2\ang{v\delta}^2+\ang{v\delta v\delta}_c$, where the subscript $c$ denotes the irreducible connected term. The momentum power spectrum, denoted by $P_q$, is typically computed by decomposing the Fourier transform, ${\bf \tilde{q}(k)}$, into components parallel to ${\bf \hat{k}}$, ${\bf \tilde{q}_\parallel = \hat{k}(\tilde{q}\cdot \hat{k}})$, and perpendicular to ${\bf \hat{k}}$, $ {\bf \tilde{q}_\perp =\tilde{q}- {\bf \hat{k}}(\tilde{q}\cdot\hat{k}})$ \cite{Ma:2002aa}. The longitudinal momentum component does not contribute significantly to $C_\ell^{TT, \text{kSZ}}$ due to cancellations of positive and negative contributions in the line-of-sight integration. For instance, as shown by Park et al. \cite{Park:2016aa}, the longitudinal contribution to $C_\ell^{TT, \text{kSZ}}$  is more than four orders of magnitude below the transverse contribution for $\ell > 3000$, so it suffices to only consider  $P_{q_\perp}$ in this calculation. \\

To compute $P_{q_\perp}$, we use the ``standard kSZ model" \cite{Park:2016aa,Hu:2000aa}, which incorporates the fully non-linear power spectrum for the density field, $P_\delta^\text{nl}$, but approximates the velocity power spectrum by linear theory $P_v = \frac{\dot{a}^2f^2}{k^2}P_\delta^\text{lin}$ where $f=\left(1+ \frac{a}{D_\Psi}\frac{dD_\Psi}{da}\right)^2$. The resulting expression is
\begin{align}
	P_{q_\perp}(k,\chi) = & \dot{a}^2 f^2 \int_{-1}^1 d\mu \int \frac{dk'}{(2\pi)^2} \ P_\delta^\text{lin}(k',\chi) \ P_\delta^\text{nl}(\sqrt{k^2-2kk'\mu+k'^2},\chi) \no \\
	& \times \frac{(k^2-2kk'\mu)(1-\mu^2)}{k^2-2kk'\mu+k'^2}. 
	\label{eq:Pqperp}
\end{align}
This model neglects the velocity-density cross correlation because the geometrical factor attached to this term decreases rapidly at large $k$. Since the kSZ contribution to $C_\ell^{TT}$ that we are interested is sensitive only to high $k$ scales, this approximation is valid. Note that this also implies our precise choice of $k_*$ in section~\ref{sec:signal} is irrelevant for the noise calculation. We are also neglecting the non-Gaussian contribution from the connected 4-point function (unimportant on all scales according to \cite{Ma:2002aa}, but could account for up to $10\%$ of the power spectrum according to \cite{Park:2016aa}). In terms of $P_{q_\perp}$ the temperature power spectrum can be expressed as
\begin{align}
	C_\ell^{TT, \text{kSZ}} = & \int dk \frac{2k^2}{\pi} \int d\chi_e \sigma_T a(\chi_e) \bar{n}_e(\chi_e) \sqrt{P_{q_\perp}(k,\chi_e)}j_\ell(k\chi_e)  \no \\
	& \times \int d\chi'_e \sigma_T a(\chi'_e) \bar{n}_e(\chi'_e) \sqrt{P_{q_\perp}(k,\chi'_e)}j_\ell(k\chi'_e) \no \\
	\simeq & \int \frac{dk}{\ell+1/2} \left[ \sigma_T a(\chi) \bar{n}_e(\chi) \right]^2 P_{q_\perp}(k,\chi) \Big|_{\chi \rightarrow (\ell+1/2)/k}, \label{eq:ClTT}
\end{align}
using the Limber approximation \cite{Loverde:2008aa} again. The functions $P_{q_\perp}(k,\chi)$ and $C_\ell^{TT, \text{kSZ}}$  are shown in figure~\ref{fig:ClTTfull}. The total temperature power spectrum is a sum of the primary and kSZ contributions, denoted from now on by 
\begin{equation}
	C_\ell^{TT} =  C_\ell^{TT,\text{p}} +  C_\ell^{TT,\text{kSZ}}.
\end{equation}

\begin{figure}[htbp]
  \begin{center}
			\subfigure{\includegraphics[width=7.5cm]{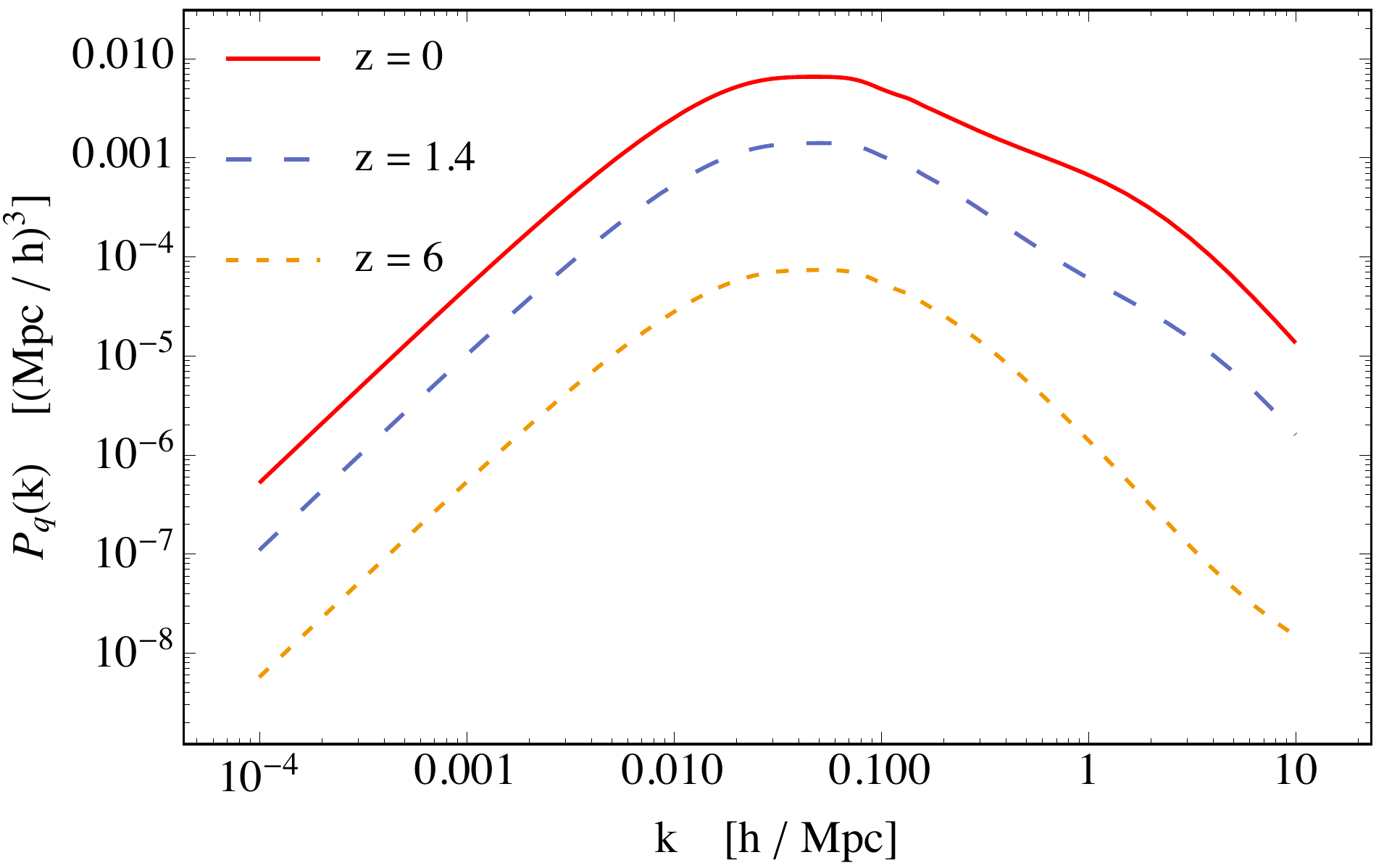}}
			\hspace{.2cm}
			\subfigure{\includegraphics[width=7.5cm]{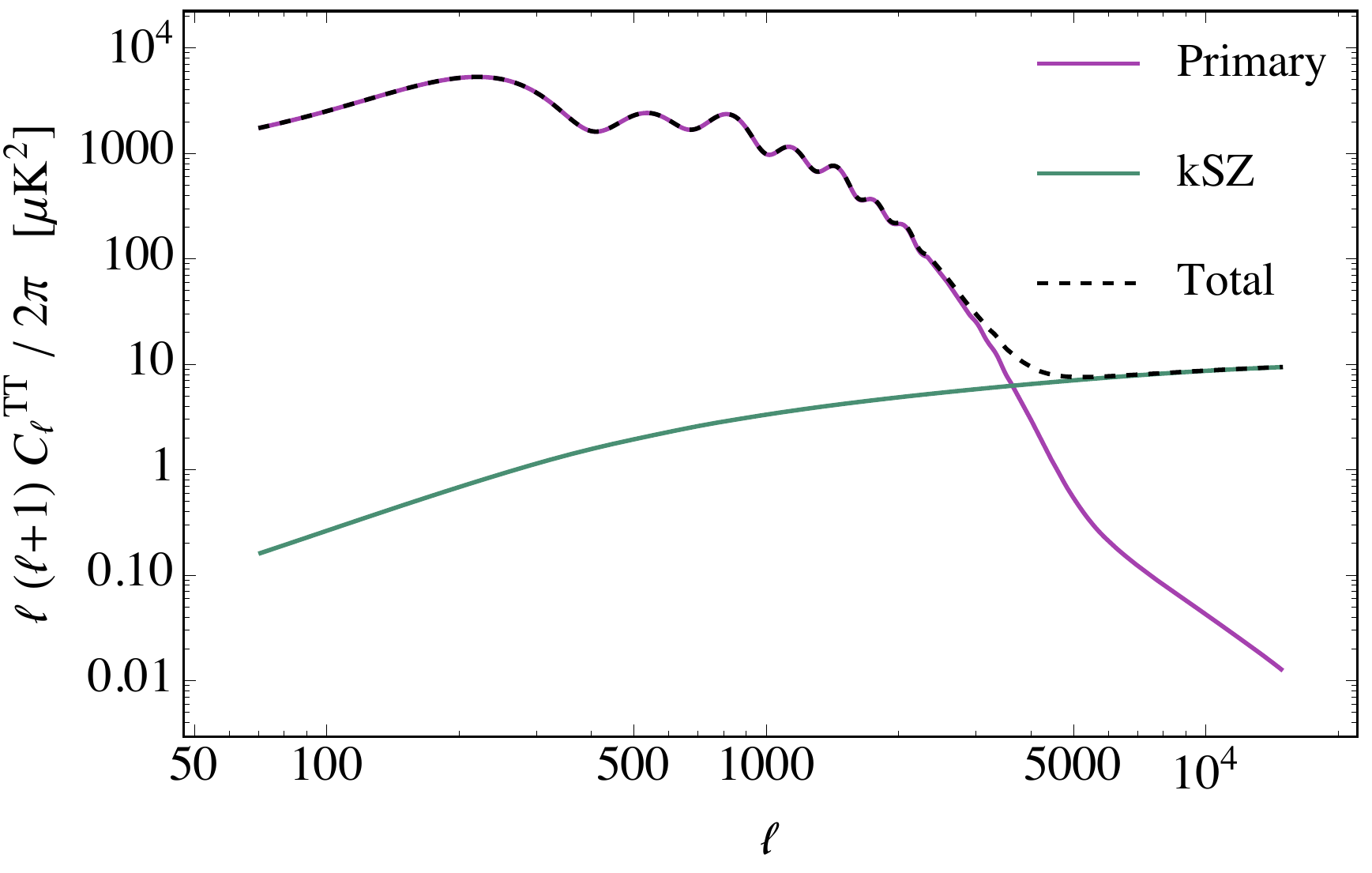}}
			\caption{Left: the power spectrum of the transverse momentum field \eqref{eq:Pqperp}. Right: the contributions to the temperature-temperature power spectrum from the primary CMB and the non-linear kSZ effect \eqref{eq:ClTT}.}
			 \label{fig:ClTTfull}
		\end{center} 
\end{figure}

The second term in eq.~\eqref{eq:4point} requires us to compute the cross correlation of the temperature with density. This can be done as follows
\begin{align}
	\bang{ \frac{\Delta T}{T}\bigg|_\text{kSZ}({\bf \hat{n}}_e)\delta({\bf \hat{n}}'_e, \bar{\chi}_e)} = & -\int d\chi_e \sigma_T a(\chi_e) \bar{n}_e(\chi_e) \ang{v_\text{eff}({\bf \hat{n}}_e,\chi_e)(1+\delta({\bf \hat{n}}_e,\chi_e)) \delta({\bf \hat{n}}'_e, \bar{\chi}_e)} \no \\
	= & -\int d\chi_e \sigma_T a(\chi_e) \bar{n}_e(\chi_e)  \int d\chi'_e W(\chi'_e, \bar{\chi}_e) \  \ang{v_\text{eff}({\bf \hat{n}}_e,\chi_e)\delta({\bf \hat{n}}'_e,\chi'_e)} \no \\
	= &  \sum_\ell \frac{2\ell+1}{4\pi} \ C_\ell^{T\delta} \  \mathcal{P}_\ell({\bf \hat{n}}_e \cdot {\bf \hat{n}}'_e).
\end{align}
Assuming approximate Gaussianity, we neglect the three point function $\ang{v\delta\delta}$, so only the correlation between $v_\text{eff}$ and $\delta$ remains. Recall that we are computing the contribution from the small scales, whereas the signal calculation above (see eqs.~\eqref{eq:dvd},\eqref{eq:crosscorrelation_intermediate}) is only sensitive to large scales. The power spectrum $C_\ell^{T\delta}$ takes the form
\begin{align}
	C_\ell^{T\delta} = & -\int \frac{dk\ k^2}{(2\pi)^3} \int d\chi_e  \sigma_T a(\chi_e) \bar{n}_e(\chi_e) \Delta_\ell^{v}(k,\chi_e)\sqrt{P_\Psi(k)} \no \\
	 & \times \int d\chi'_e 4\pi \sqrt{P_\delta(k,\chi'_e)}W(\chi'_e, \bar{\chi}_e)j_\ell(k\chi'_e) \no \\
	= & -\int \frac{dk\ k^2}{(2\pi)^3} \int d\chi_e  \sigma_T a(\chi_e) \bar{n}_e(\chi_e) 4\pi \mathcal{K}^v(k,\chi_e)\frac{T(k)}{k}\frac{dj_\ell(k\chi_e)}{d\chi_e} \sqrt{P_\Psi(k)}  \no \\
	& \times \int d\chi'_e 4\pi \sqrt{P_\delta(k,\chi'_e)}W(\chi'_e, \bar{\chi}_e)j_\ell(k\chi'_e),
\end{align}
where we've recalled the expression for $\Delta_\ell^{v}(k,\chi_e)$ from eq.~\eqref{eq:transfer}, and used the identity,
\begin{equation}
	\frac{\ell\ j_{\ell - 1} (k \chi) - (\ell + 1)  j_{\ell + 1} (k \chi)}{2\ell+1} = \frac{1}{k}\frac{dj_\ell(k\chi)}{d\chi} .
\end{equation}
Integrating by parts and using the Limber approximation \cite{Loverde:2008aa} results in the expression
\begin{align}
	C_\ell^{T\delta} = & \int dk \frac{2k^2}{\pi} \int d\chi_e  \sigma_T \frac{d}{d\chi_e}[a(\chi_e) \bar{n}_e(\chi_e) \mathcal{K}^v(k,\chi_e)]\frac{T(k)}{k} \sqrt{P_\Psi(k)} j_\ell(k\chi_e) \no \\ 
	& \times \int d\chi'_e \sqrt{P_\delta(k,\chi'_e)}W(\chi'_e, \bar{\chi}_e)j_\ell(k\chi'_e) \\
	\simeq & \int \frac{dk}{\ell+1/2}\  \sigma_T \ \frac{d}{d\chi}[a(\chi) \bar{n}_e(\chi) \mathcal{K}^v(k,\chi)] \ \frac{T(k)}{k} \ \sqrt{P_\Psi(k)P_\delta(k,\chi)} \ W(\chi, \bar{\chi}_e) \Big|_{\chi \rightarrow (\ell+1/2)/k} . \no
\end{align}

\begin{figure}[htbp]
	\begin{center}
	\includegraphics[width=9cm]{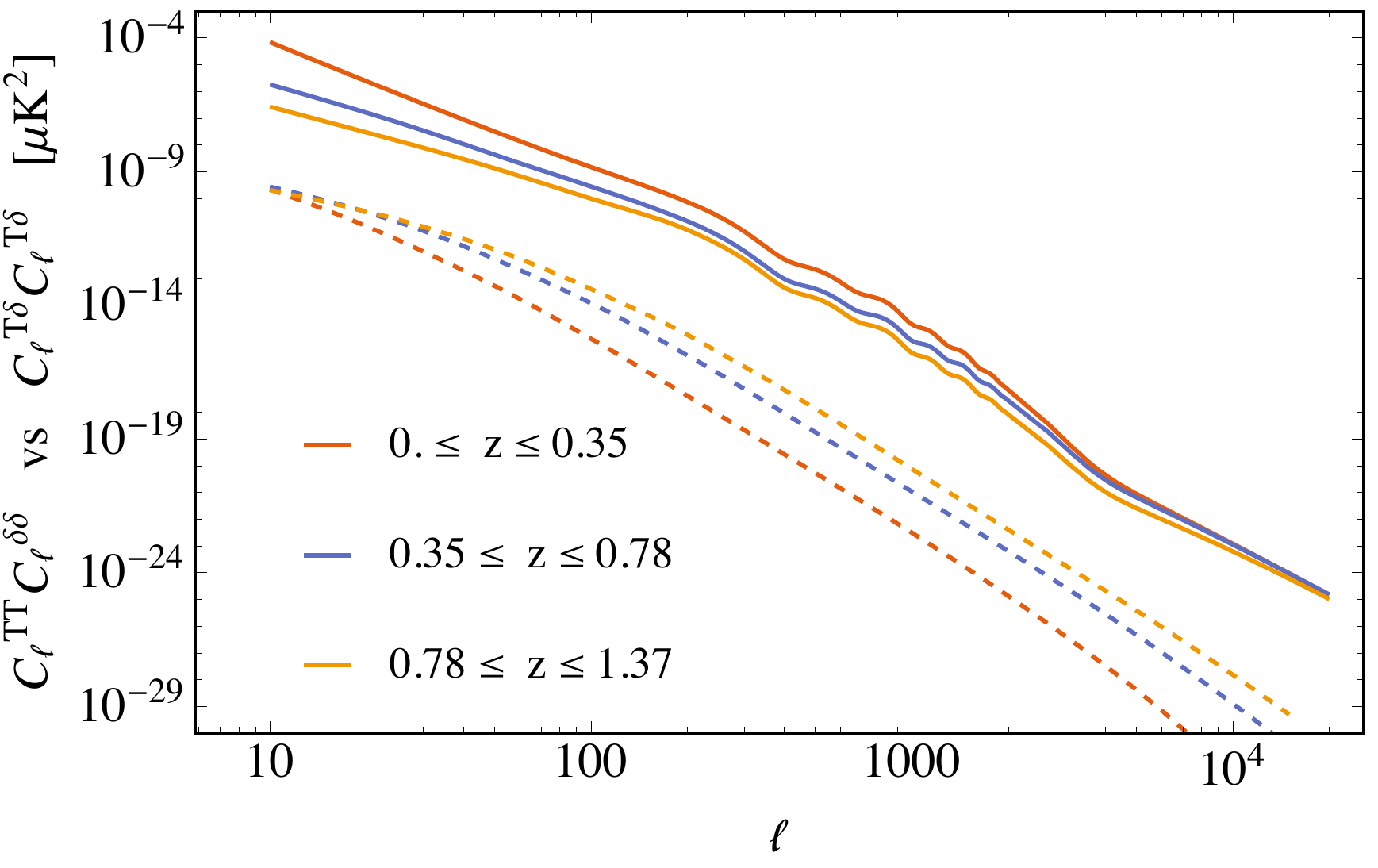}
	\caption{The solid curves are $C_\ell^{TT} C_{\ell}^{\delta\delta}$ and the dashed curves are $C_\ell^{T\delta} C_{\ell}^{T\delta}$, which represent the two terms in square brackets in eq.~\eqref{eq:noisesum} evaluated at the same $\ell$. The redshift ranges indicate sample redshift bins (see figure \ref{fig:bins}).}
	\label{fig:ClTd}
	\end{center}
\end{figure}

Combining all of these pieces, expanding $\mathcal{P}_\ell({\bf \hat{n}}_e \cdot {\bf \hat{n}}'_e)$ in terms of spherical harmonics using the identity \eqref{eq:PYY}, the variance in the power multipoles can then be written
\begin{align}
	\ang{\tilde{b}_L^2}= & \sum_M\frac{\ang{\tilde{b}^*_{LM} \tilde{b}_{LM}}}{2L+1} \no \\
	= & \sum_{\ell,\ell',m,m',M}\frac{C_\ell^{TT}C_{\ell'}^{\delta\delta}+C_\ell^{T\delta}C_{\ell'}^{T\delta}}{2L+1}\ \Big| \int d^2{\bf \hat{n}}_e Y_{LM}({\bf \hat{n}}_e) Y^*_{\ell m }({\bf \hat{n}}_e) Y^*_{\ell' m' }({\bf \hat{n}}_e) \Big|^2 \no \\
	= & \sum_{\ell,\ell'}[C_\ell^{TT}C_{\ell'}^{\delta\delta} +C_\ell^{T\delta}C_{\ell'}^{T\delta}]\ \frac{(2\ell+1)(2\ell'+1)}{4\pi(2L+1)^2} \ \Big| \text{C}^{\ell \ell' L}_{0 0 0} \Big|^2 \sum_{m,m',M} \Big| \text{C}^{\ell \ell' L}_{m m' M} \Big|^2 \no \\
	= & \sum_{\ell,\ell'}[C_\ell^{TT}C_{\ell'}^{\delta\delta} +C_\ell^{T\delta}C_{\ell'}^{T\delta}]\ \frac{(2\ell+1)(2\ell'+1)}{4\pi(2L+1)} \ \Big| \text{C}^{\ell \ell' L}_{0 0 0} \Big|^2 ,\label{eq:noisesum}
\end{align}
where $\text{C}^{\ell \ell' L}_{m m' M}$ are the Clebsch-Gordan coefficients, and we used the triple product integral identity (see eq.~\eqref{eq:YYY}) and $\sum_{m,m',M} \Big| \text{C}^{\ell \ell' L}_{m m' M} \Big|^2 = 2L+1$. Note that the coefficients $\text{C}^{\ell \ell' L}_{0 0 0}$ are only nonzero for $|\ell-\ell'| \leq L \leq \ell+\ell'$ and $\ell+\ell'+L$ even. Figure~\ref{fig:ClTd} shows $(C_\ell^{T\delta})^2$ plotted in three different redshift bins, and compares this to $C_\ell^{TT}C_\ell^{\delta\delta}$. It is clear that the cross term is subdominant by several orders of magnitude for all $\ell$, and therefore it is sufficient to approximate $C_\ell^{TT} C_{\ell'}^{\delta\delta}+C_\ell^{T\delta} C_{\ell'}^{T\delta} \sim C_\ell^{TT} C_{\ell'}^{\delta\delta}$.

The final result for the cosmic variance limited noise estimate is
\begin{equation} \label{eq:noise}
\sqrt{\ang{\tilde{b}_L (\bar{\chi}_e)^2}} = \sqrt{\sum_{\ell,\ell'=\ell_\text{min}}^{\ell_\text{max}}[C_\ell^{TT}C_{\ell'}^{\delta\delta} (\bar{\chi}_e)]\ \frac{(2\ell+1)(2\ell'+1)}{4\pi(2L+1)} \ \Big| \text{C}^{\ell \ell' L}_{0 0 0} \Big|^2} .
\end{equation}
where we have restored an explicit dependence on the redshift bin $\bar{\chi}_e$. Notice that the result is given by a sum over $\ell,\ell'$, and recall that the noise captures the accidental power asymmetry which is sensitive only to high $\ell$. In the spirit of our cosmic variance limited estimate, we can imagine that we have sufficiently filtered out the primary CMB and cleaned the foregrounds. We capture this in the calculation by starting the sum in \eqref{eq:noise} at $\ell_\text{min}=3000$. As previously mentioned, we will employ the default values  ($\ell_\text{min}=3000,\ \ell_\text{max}=\infty$) unless otherwise stated. Results for varying $\ell_\text{min}$ and $\ell_\text{max}$ will be presented in section~\ref{sec:detectability}, allowing us to make statements about the detectability with next generation CMB experiments and galaxy surveys.

\section{Cosmic Variance Limited Signal to Noise} \label{sec:estsig}

In this section, we assess the signal to noise for the power multipoles eq.~\eqref{eq:sigreal} in the cosmic variance limit using both a theoretical estimate and the simulations described in section~\ref{sec:simulations}. In each case, the signal to noise in each bin $\bar{\chi}_e$ is calculated as
\begin{equation} \label{eq:SN}
	\frac{S}{N} (\bar{\chi}_e)= \frac{b_{LM} (\bar{\chi}_e)}{\sqrt{\ang{\tilde{b}_L (\bar{\chi}_e)^2}}} 
\end{equation}
where $b_{LM} (\bar{\chi}_e)$ is found using eq.~\eqref{eq:sigreal} and $\sqrt{\ang{\tilde{b}_L (\bar{\chi}_e)^2}}$ is given by eq.~\eqref{eq:noise}.

\subsection{RMS Estimate}

A simple estimate of the signal is obtained by approximating $a_{LM}^v(\chi_e)\sim\sqrt{C_L^v(\chi_e)}$ in eq.~\eqref{eq:sigreal}, where $C_L^v(\chi_e)$ is the power spectrum associated with the large-scale velocity, given by
\begin{equation} \label{eq:clv}
	C_L^v(\chi_e) = \int_0^{k_\text{max}} \frac{k^2 dk}{(2\pi)^3}P_\Psi(k) |\Delta^v_{L}(k, \chi_e)|^2
\end{equation}
with $\Delta^v_{L}(k, \chi_e)$ given by eq.~\eqref{eq:transfer}. This necessarily yields an overestimate of the signal, since in reality $a_{L0}^v(\chi_e)$ will vary over over the window functions leading to partial cancellation, while here we are assuming that it always takes its (positive definite) RMS value. By comparing with simulations in the following subsection, we show that this approximation gives a good estimate in the limit of thin window functions $W(\chi_e, \bar{\chi}_e)$ (in the context of this paper, this is equivalent to the limit of many redshift bins). 

\begin{figure}[htbp]
	\begin{center}
	\includegraphics[width=9cm]{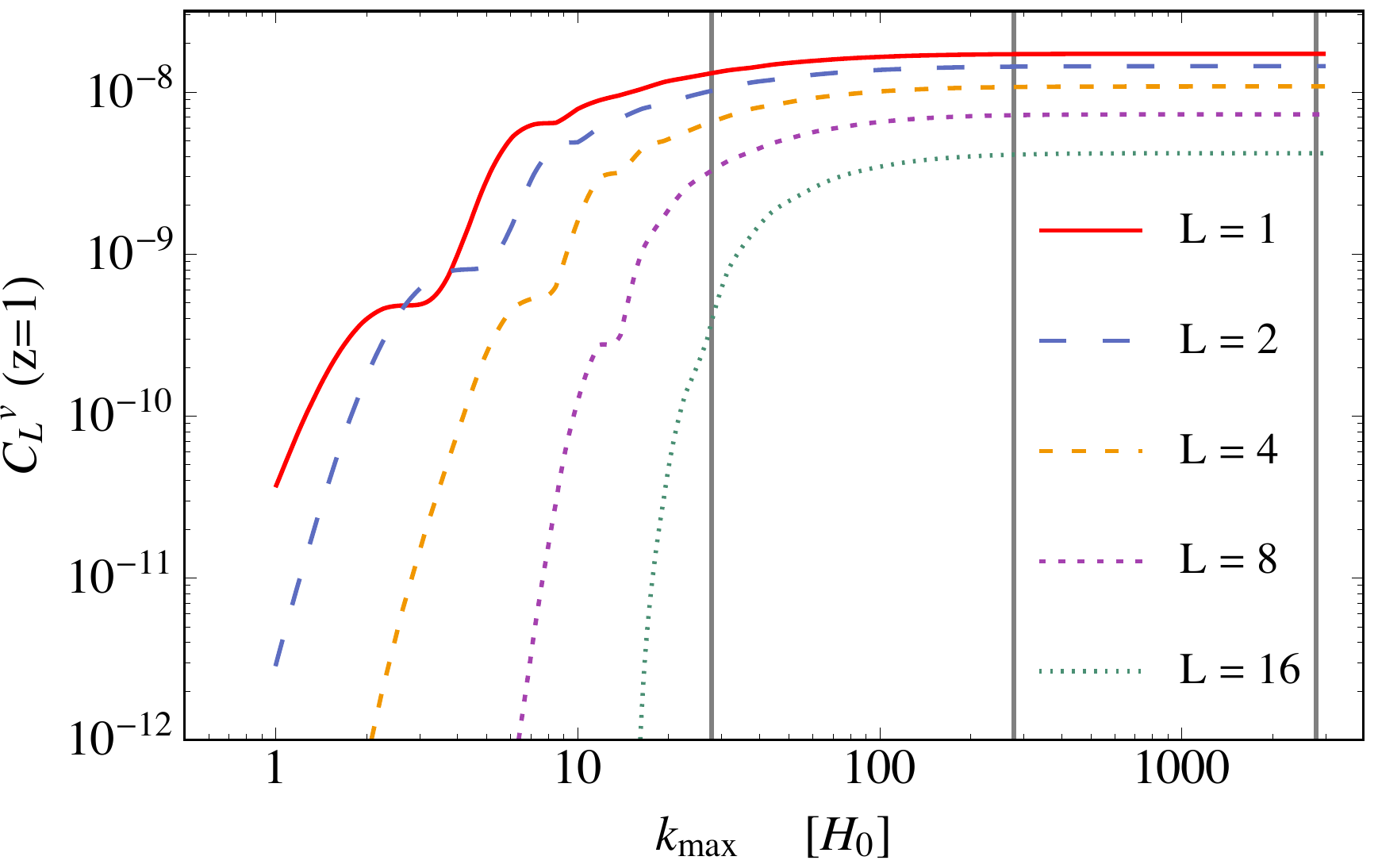}
	\caption{The effective velocity contribution to the signal described in eq.~\eqref{eq:clv} versus $k_\text{max}$ in units of $H_0$, which refers to the upper bound on the integral. Notice that the contribution mainly comes from large scales (small $k$) for low $L$. The vertical lines indicate scales equal to 1 Gpc, 100 Mpc, and 10 Mpc from left to right.}
	\label{fig:clv}
	\end{center}
\end{figure}

Before proceeding, we can assess which scales form the dominant contribution to eq.~\eqref{eq:clv}. The upper limit of integration in eq.~\eqref{eq:clv}, $k_\text{max}$, corresponds to the smallest scale, $\lambda_\text{min}=2\pi/k_\text{max}$, that contributes to the signal. Formally, $k_\text{max} \rightarrow \infty$, but we can adjust the cutoff to include only the long modes discussed above. In figure~\ref{fig:clv} we show $C_L^v$ at $z=1$ as a function of $k_{\rm max}$. Vertical lines indicate $\lambda_\text{min} = 10, 10^2, 10^3 \ {\rm Mpc}$. Here, we see that for a number of power multipoles, the relevant signal is obtained almost entirely from scales $\sim 10^2 - 10^3 \ {\rm Mpc}$.

Putting everything together, applying the Limber approximation \cite{Loverde:2008aa}, and choosing $M=0$ under the assumption that all other $M$ will statistically be the same, the final expression for the signal becomes
\begin{equation} \label{eq:signal}
	b_{L0} (\bar{\chi}_e) 
	 \simeq \sum_{\ell=\ell_\text{min}}^{\ell_\text{max}} \frac{2\ell+1}{4\pi} \int \frac{d\chi}{\chi^2} \ \sigma_T \ a(\chi) \ \bar{n}_e(\chi) \ \sqrt{C_L^v(\chi)} \ W(\chi, \bar{\chi}_e) \ P_\delta\left(\frac{\ell+1/2}{\chi},\chi\right).
\end{equation}

\begin{figure}[htbp]
	\begin{center}
	\includegraphics[width=15.5cm]{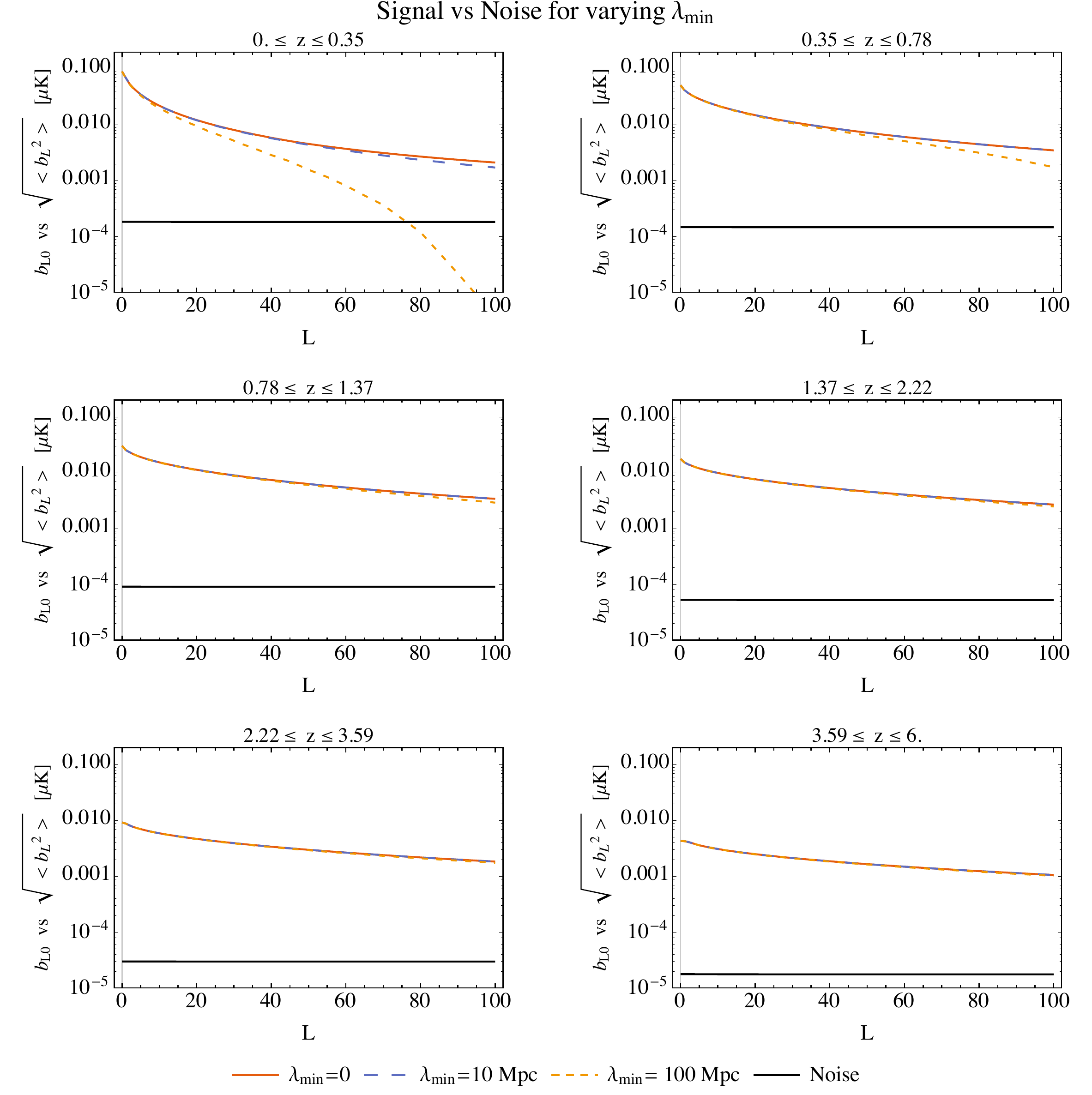}
	\caption{The signal \eqref{eq:signal} is shown in six redshift bins, calculated for three different scales: $k_\text{max} = 278\ H_0$ ($\lambda_\text{min} = 100$ Mpc) [orange, short dash], $k_\text{max} = 2780\ H_0$ ($\lambda_\text{min} = 10$ Mpc) [blue, long dash], and $k_\text{max} = \infty$ ($\lambda_\text{min} = 0$) [red, solid]. The solid black line is the noise estimate $\sqrt{\ang{\tilde{b}_L^2}}$ (eq. \eqref{eq:noise}), which falls well below the estimated signal for this configuration. The summation bounds employed for the signal and noise calculations are ($\ell_\text{min}=3000,\ \ell_\text{max}=\infty$).}
	\label{fig:sigs}
	\end{center}
\end{figure}

We compute the signal eq.~\eqref{eq:signal} and noise eq.~\eqref{eq:noise} for the $N_\text{bin}=6, \ 12, \ \text{and} \ 24$ top-hat bin configurations with redshift ranges summarized in figure~\ref{fig:bins}. We show the results for the 6-bin configuration in figure~\ref{fig:sigs}. Each plot has four curves. The solid red curve is the RMS signal with $k_\text{max} \rightarrow \infty$, the dotted orange line is the cosmic variance limited noise, the short dashed blue line is the RMS signal computed for $k_\text{max} = 278\ H_0$ ($\lambda_\text{min} = 100$ Mpc), and the long dashed green curve is the signal computed for $k_\text{max} = 2780\ H_0$ ($\lambda_\text{min} = 10$ Mpc). Comparing the three signal curves we see that except in the lowest redshift bin at high $L$, the signal is composed primarily of long wavelength modes ($\lambda > 100$ Mpc), as expected. The amplitude of the signal varies by roughly an order of magnitude between the lowest and highest redshift bin, and is strongest at low $L$ and low redshift. Both the signal and noise are relatively flat over the plotted range in $L$. Most importantly though, the signal is 2-3 orders of magnitude larger than the noise!

\begin{figure}[htbp]
	\begin{center}
	\includegraphics[width=9cm]{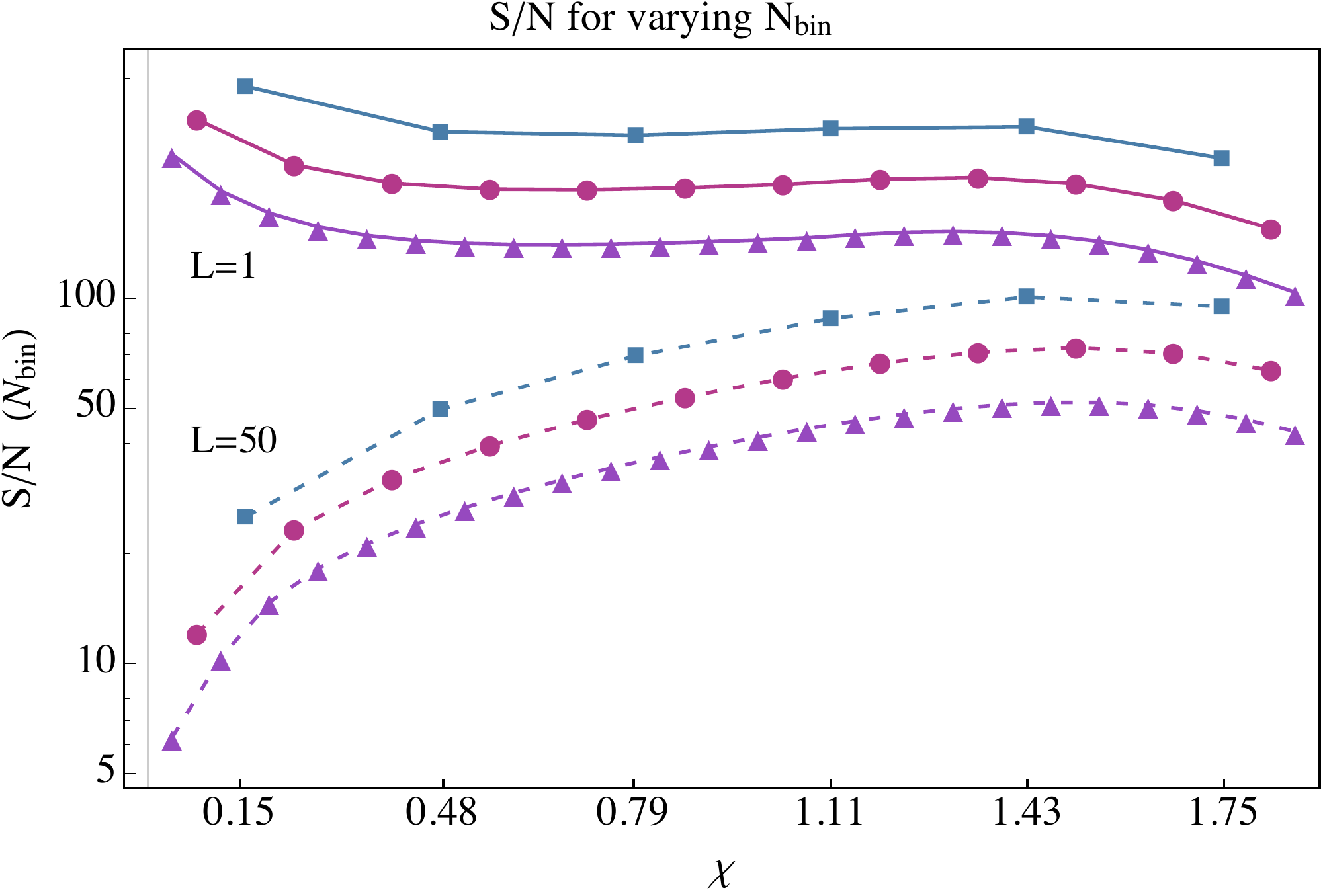}
	\caption{The bullet points (square, circle, triangle) represent the signal-to-noise (eq. \eqref{eq:SN}) in each bin for $N_\text{bin}= (6,12,24)$, plotted against $\chi$ at the midpoint of the bin. The solid top curves are for $L=1$ and the bottom dashed curves are for $L=50$. The signal and noise were computed using eq. \eqref{eq:signal} and \eqref{eq:noise}, employing summation bounds ($\ell_\text{min}=3000,\ \ell_\text{max}=\infty$), and with $k_\text{max} \rightarrow \infty$ in eq. \eqref{eq:clv}.}
	\label{fig:6vs12}
	\end{center}
\end{figure}

To compare the result for the $12$ and $24$ bin configurations, in figure~\ref{fig:6vs12} we show the signal to noise eq.~\eqref{eq:SN} at $L=1$ and $L=50$ as a function of comoving distance. Increasing the number of bins by a factor of 2 results in a decrease in the signal-to-noise by a factor of $\sim \sqrt{2}$. This is true for all $L$. We therefore conclude that a signal can in principle be measured in the cosmic variance limit at high signal to noise for a large number of redshift bins $N_{\rm bins}$ at a variety of scales $L$.

\subsection{Comparing with the signal from simulations} \label{sec:realizations}

The above calculation is an over-estimate of the signal, as it assumes the velocity field is positive definite and given by its RMS value. In particular, it does not account for partial cancellations along the line of sight. To take this into account, we can compute the signal from eq.~\eqref{eq:sigreal} using the effective velocity field computed from the simulations described in section~\ref{sec:simulations}. Using an interpolating function for $a_{\ell m}^{v} (\chi_e)$ constructed from the $v_{\rm eff}$ map in each of the 50 redshift bins, we compute eq.~\eqref{eq:sigreal} directly for 100 realizations at a resolution of  $k_\text{max}\sim57.4\ H_0$ for the 6 and 12 bin configurations. Below we only present results for $b_{L0}$; other values of $M$ have identical statistical properties.

\begin{figure}[htbp]
	\begin{center}
	\includegraphics[width=15.5cm]{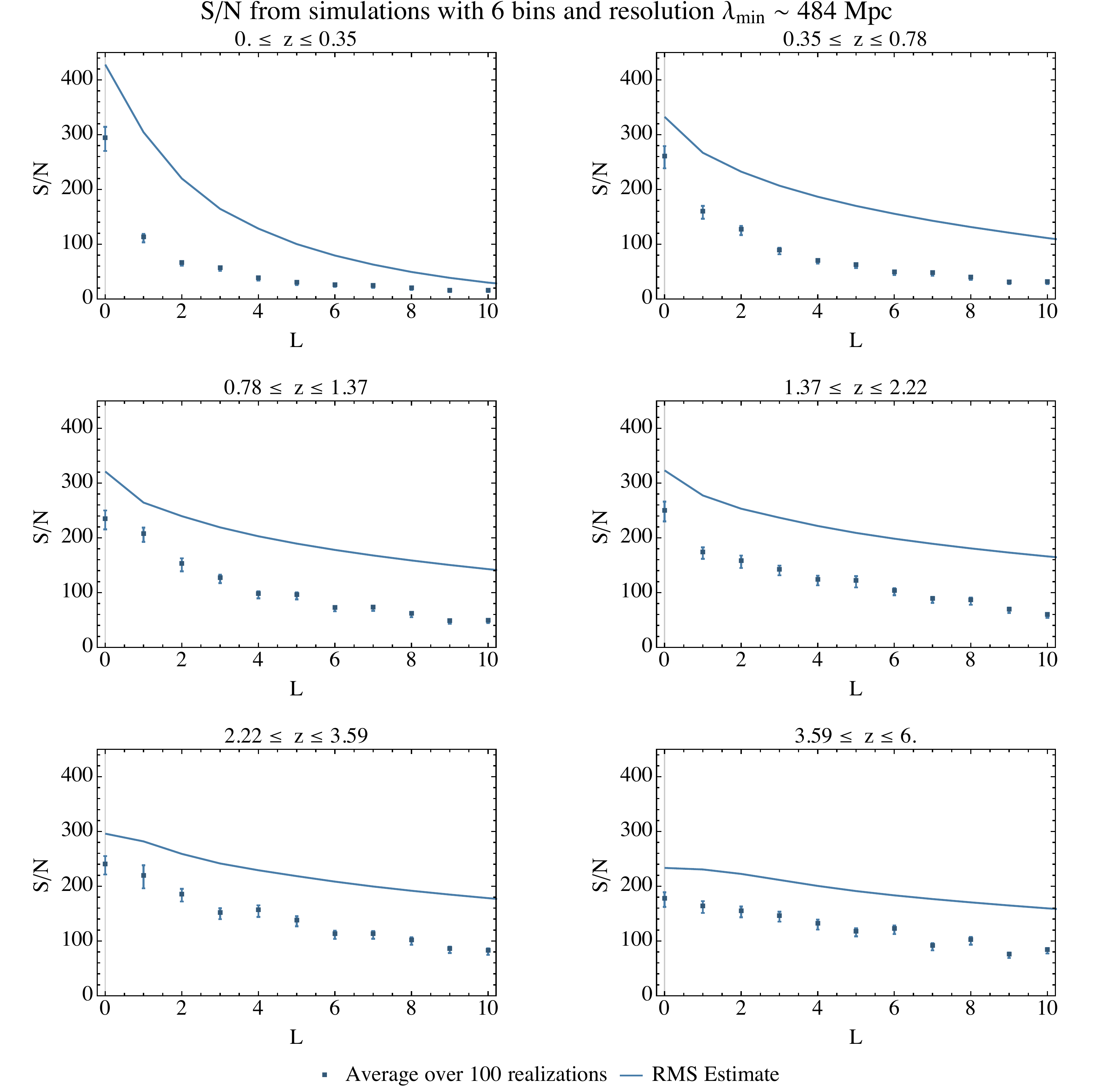}
	\caption{The signal-to-noise (eq. \eqref{eq:SN}) computed using simulations, in comparison to the RMS estimate (solid curves), in 6 redshift bins (see figure~\ref{fig:bins}). The points represent the standard deviation of the 100 realizations, and the error bars denote the standard error of the standard deviation. These simulations have a resolution of $\lambda_\text{min} \sim 484$~Mpc and utilized the default summation bounds ($\ell_\text{min}=3000,\ \ell_\text{max}=\infty$) for the signal and noise calculations.}
	\label{fig:SN6}
	\end{center}
\end{figure}

\begin{figure}[htbp]
	\begin{center}
	\includegraphics[width=16cm]{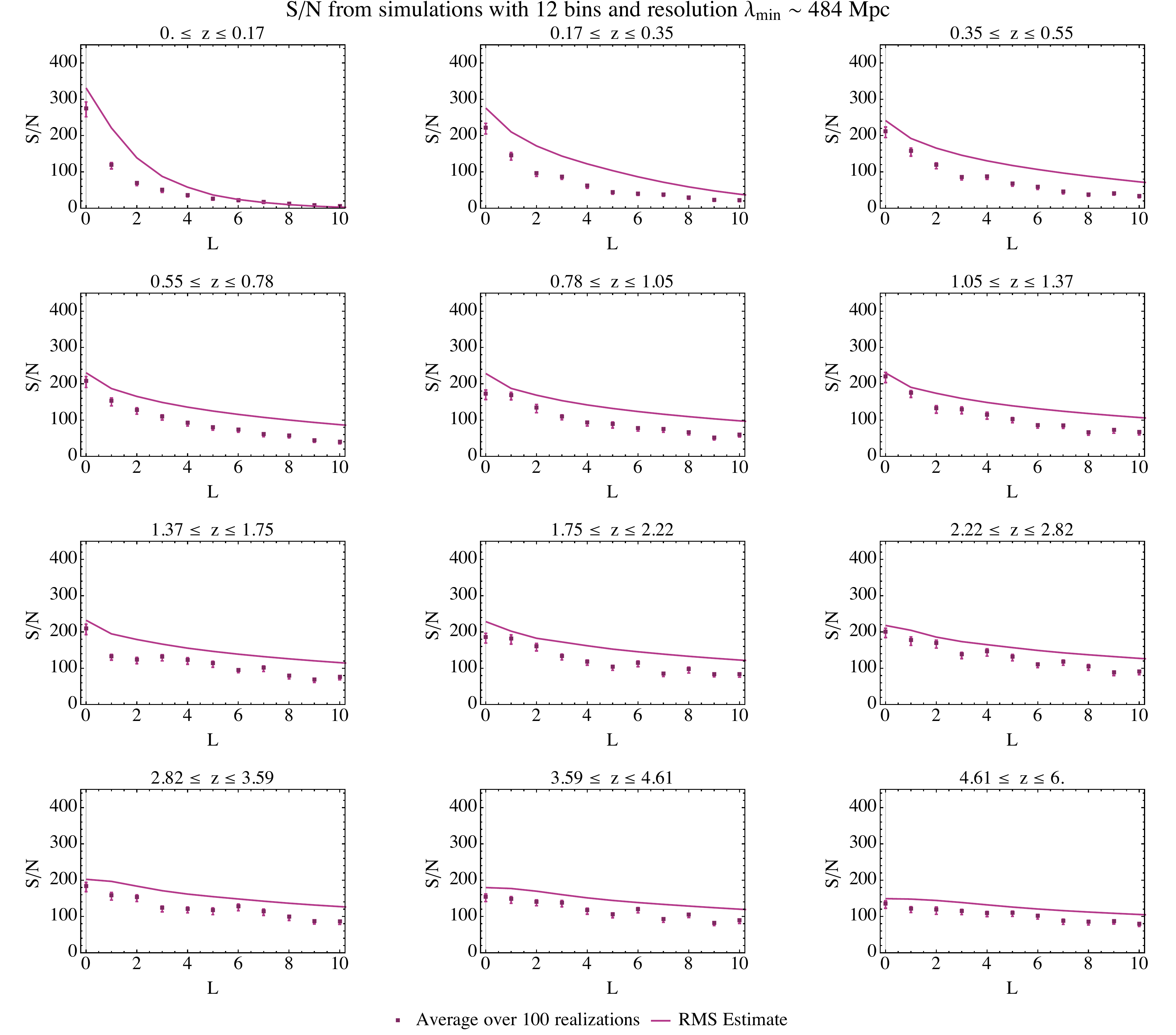}
	\caption{[The same as Figure~\ref{fig:SN6} but for 12 redshift bins]. The signal-to-noise (eq. \eqref{eq:SN}) computed using simulations, in comparison to the RMS estimate (solid curves), in 12 redshift bins (see figure~\ref{fig:bins}). The points represent the standard deviation of the 100 realizations, and the error bars denote the standard error of the standard deviation. These simulations have a resolution of $\lambda_\text{min} \sim 484$~Mpc and utilized the default summation bounds ($\ell_\text{min}=3000,\ \ell_\text{max}=\infty$) for the signal and noise calculations.}
	\label{fig:SN12}
	\end{center}
\end{figure}

Figures \ref{fig:SN6} and \ref{fig:SN12} shows the signal to noise computed using simulations, in comparison to the one estimated from theory with eq.~\eqref{eq:clv} computed using the integration limit $k_\text{max}\sim57.4\ H_0$ corresponding to the simulation resolution. We plot $L<10$, which is accurately captured for the resolution we consider (see figure~\ref{fig:clv}). Notice that the solid curves, showing the predicted signal based on our theory calculation, are always higher than the average signal from the realizations. This is due to a difference in the order of operations. In our estimation in \eqref{eq:signal} we have averaged over $a_{L0}(\chi_e)$ prior to integrating over $\chi_e$, whereas the signal computed from the realizations integrates over $\chi_e$ first and then averages. Since $a_{L0}(\chi_e)$ is an oscillating function that takes positive and negative values, there can be cancellation upon integration over $\chi_e$. This cancellation can be minimized by using smaller redshift bins, which results in a smaller range of integration and a lesser chance for cancellation. This can be noticed empirically as the agreement between realizations and theory is better for the 12 bin configuration than the 6 bin. In summary, the realizations approach the theory estimation for more bins as a result of having less variation in $a_{LM}$ over the bin. \\

\section{Mode counting} \label{sec:modes}

With an understanding of this signal in hand, we now want to estimate how many modes one could conceivably measure in the cosmic variance limit. Let's consider scales $\lambda > 100 $ Mpc, of order the BAO scale. This corresponds to $k_\text{max} = 278\ H_0$ and $L_\text{max} \sim \pi/\theta \sim \pi\chi_\text{dec}/\lambda_\text{min} = k_\text{max}\chi_\text{dec}/2 \sim 437$. On this scale and larger, the primary CMB contains $\sum_{L}^{437}({2L+1})=191843$ modes.\\

For the kSZ effect considered here, the sum over $2L+1$ is performed in each bin, therefore one might naively guess that it is possible to get $N_\text{bin}$ times more modes than the primary CMB. However, a more careful estimate needs to be done because the value of $L_\text{max}$ varies in each bin depending on the size of the signal. Consider the signal-to-noise, described by the ratio of \eqref{eq:signal} and \eqref{eq:noise}, for the scales considered here (see eq. \eqref{eq:clv} and use $k_\text{max} = 278\ H_0$). For $N_\text{bin}$ bins, in each bin, the value of $L_\text{max}$ up to which modes can be measured is found by ensuring that 
\begin{enumerate}
	\item For $L<L_\text{max}$, the signal-to-noise is bigger than 1.
	\item For $L<L_\text{max}$, the signal is dominated by modes larger than $100$ Mpc. More precisely, $b_{L0}^{k_\text{max}} > 0.95 b_{L0}^{\infty}$ for $k_\text{max} = 278\ H_0$.\footnote[3]{The $95\%$ criteria is arbitrary.}
\end{enumerate}
Table \ref{tab:modes} shows the values of $L_\text{max}$ computed in each bin for the 6, 12 and 24 bin configurations. In every case, it was a failure of criteria (2.) that determined $L_\text{max}$, as the signal-to-noise is always much bigger than 1 for this range in $L$. Notice that by doubling the bin size, we approximately double the number of modes. This allows us to extrapolate our results from the three bin configurations.

\begin{table}[]
\centering
\begin{tabular}{|c|p{8cm}|c|}
\hline
\multicolumn{1}{|c|}{$N_\text{bin}$} & \multicolumn{1}{c|}{$L_\text{max}$} & \multicolumn{1}{c|}{$\sum\limits_\text{bin}\sum\limits_{L}^{L_\text{max}}({2L+1})$} \\ \hline
6 & 6, 32, 55, 78, 100, 123 & 36086 \\ \hline
12 & 4, 16, 28, 39, 51, 62, 73, 85, 96, 107, 118, 131 & 74946 \\ \hline
24 & 3, 8, 14, 20, 25, 31, 37, 42, 48, 54, 59, 65, 71, 76, 82, 88, 93, 99, \
105, 110, 116, 121, 130, 132 & 150853 \\ \hline
\end{tabular}
\caption{Number of modes }
\label{tab:modes}
\end{table}

\begin{figure}[htbp]
	\begin{center}
	\includegraphics[width=10cm]{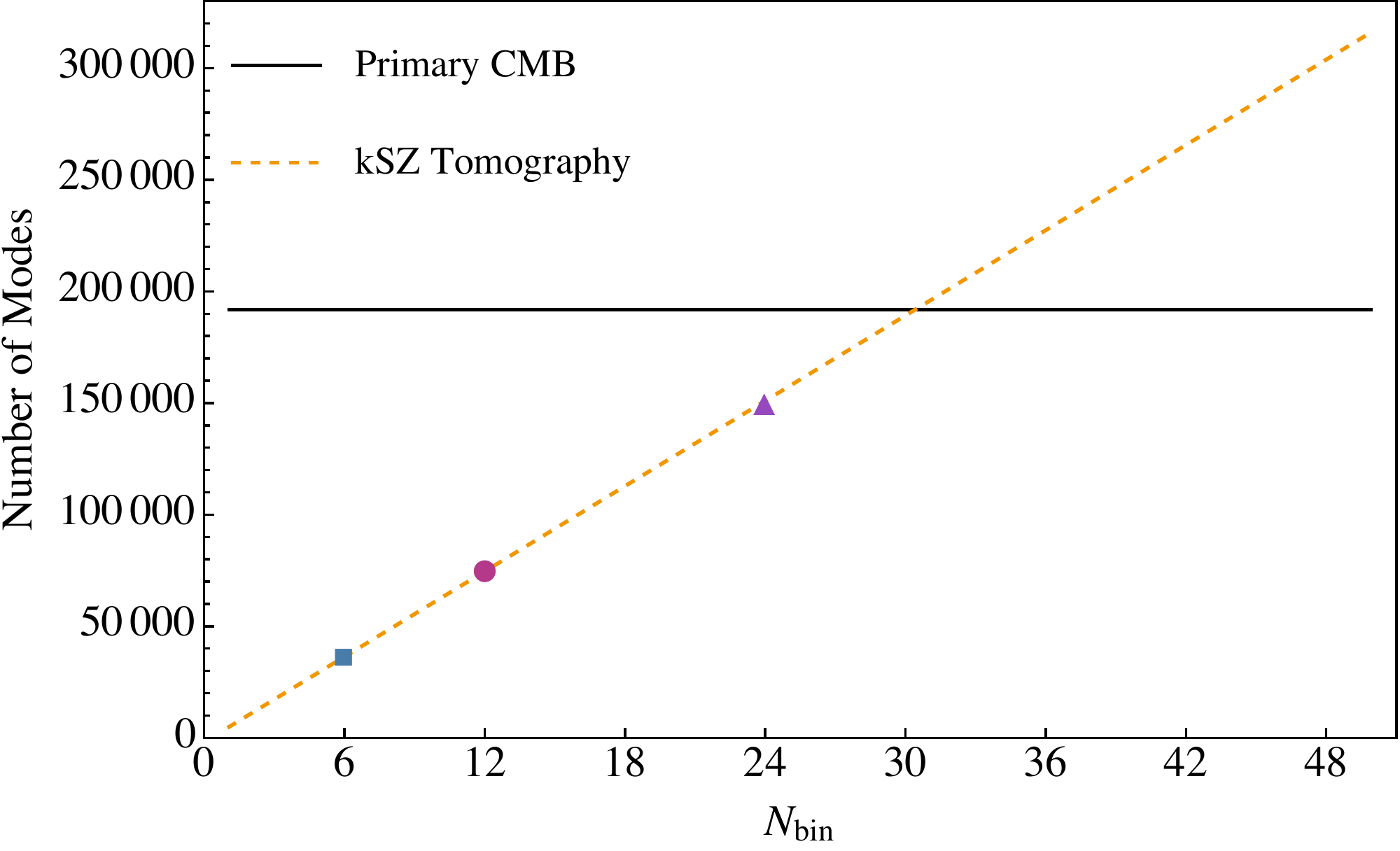}
	\caption{The bullet points show the number of modes that can possibly be measured using $N_\text{bin}= $ 6, 12, and 24. Extrapolating these data points (dashed line) indicates that we need at least 30 bins to match the number of modes in the primary CMB at $\ell \leq 437$ (solid line). These calculations were performed with the default summation bounds ($\ell_\text{min}=3000,\ \ell_\text{max}=\infty$).}
	\label{fig:modes}
	\end{center}
\end{figure}

Figure \ref{fig:modes} shows how the number of modes increases with $N_\text{bin}$ based on our estimates using 6, 12 and 24 bins. Extrapolating the data points to higher values of $N_\text{bin}$ shows that at least 30 bins are needed to match the number of modes in the primary CMB. The same increasing trend should continue until $N_\text{bin}\sim 50$ (producing 309656 modes), at which point we estimate that the signal-to-noise will drop below 1 in the high redshift bins, thus failing to satisfy criteria (1.), and causing a less rapid increase in the number of modes as a function of $N_\text{bin}$. Therefore, this method can theoretically achieve more modes than the primary CMB. However, note that within the observable Universe, on scales $\lambda > 100$ Mpc, there are approximately $\left( \chi_\text{dec} k_\text{max}/(2\pi) \right)^3 \sim 2.7 \times 10^6$ total modes. Thus, while remaining competitive with the primary CMB, the proposed method still falls about an order of magnitude short of the \emph{total} number of possible modes. In addition, because they provide at least partially independent constraints, combining the information from the primary CMB and the information from kSZ tomography can in principle constrain a larger number of modes than either individually.

\section{Detectability}\label{sec:detectability}

How close can we get to the cosmic variance limited result with the next generation of CMB experiments and galaxy surveys? Although a complete treatment is beyond the scope of this paper, we can give a rough estimate here. The two parameters affecting detectability in the analysis above are the filtering scale $\ell_\text{min}$ and resolution $\ell_\text{max}$. The filtering scale is a parameter which can be optimized in any hypothetical analysis. By varying $\ell_\text{max}$, we can define a rough target for the instrumental noise, resolution, and foreground residuals necessary for a detection to be made. 


Previously, we had made the fiducial choice $\ell_\text{min} = 3000$, which is roughly the angular scale on which the kSZ power surpasses that in the primary CMB. Here, we will explore the filtering scales $\ell_\text{min} = 2, \ 1000, \ 3000$. In addition, we chose $\ell_\text{max} \rightarrow \infty$, corresponding to the cosmic variance limit. Here, we consider a low-resolution scenario with $\ell_\text{max} = 3000$ and high-resolution scenario with $\ell_\text{max} = 5000$. 

\begin{figure}[htbp]
	\begin{center}
	\includegraphics[width=15.5cm]{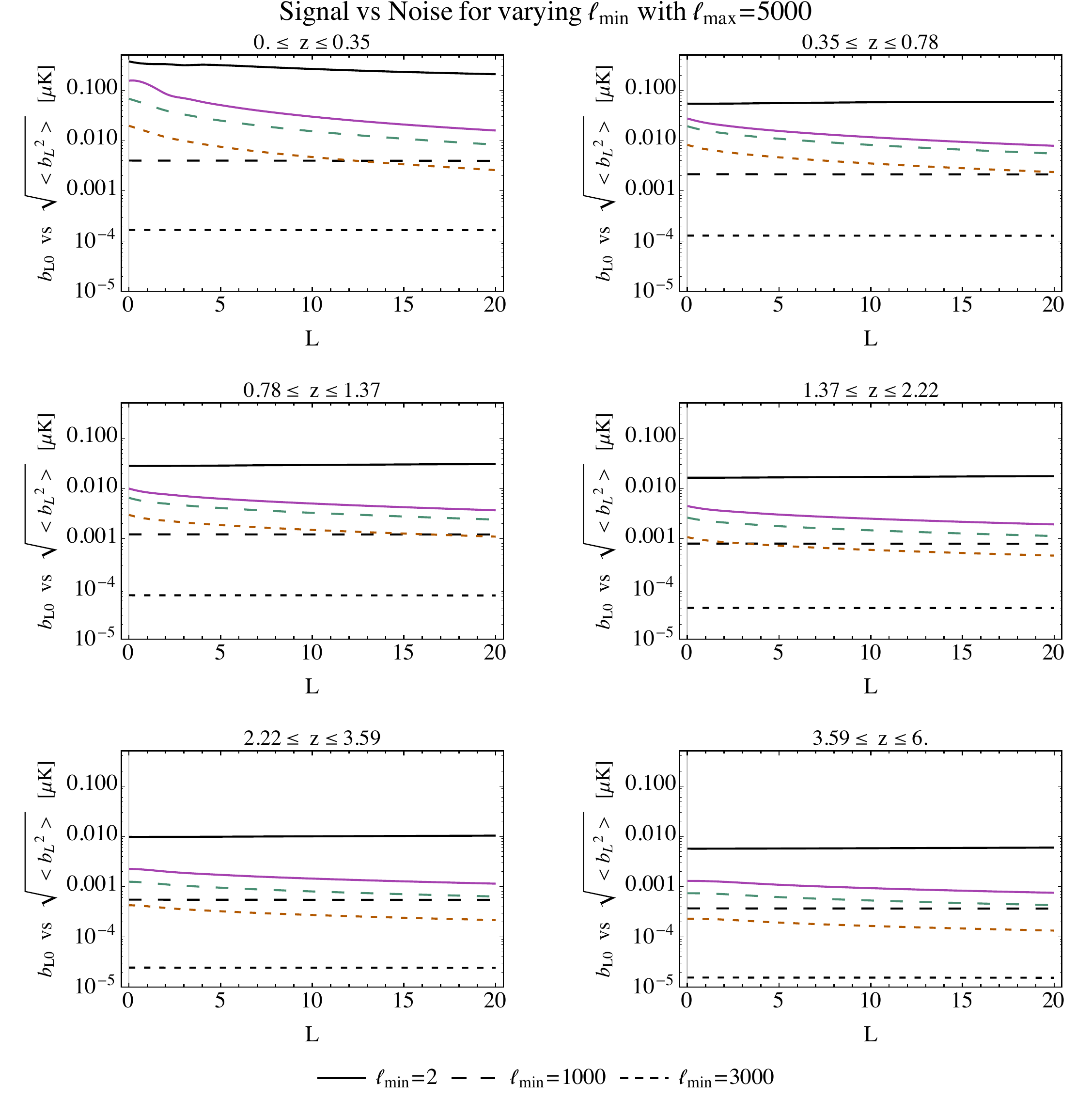}
	\caption{The signal \eqref{eq:signal} (coloured lines) and noise  \eqref{eq:noise} (black lines) are shown in six redshift bins, computed with $k_\text{max} = \infty$ and $\ell_\text{max}=5000$. We consider different filter scales: $\ell_\text{min}=2$ (solid), $\ell_\text{min}=1000$ (long dash), $\ell_\text{min}=3000$ (short dash).}
	\label{fig:lmax5000}
	\end{center}
\end{figure}

Starting with the high resolution scenario, in figure~\ref{fig:lmax5000}, we show the signal and noise for six redshift bins, choosing $\ell_\text{min} = 2, \ 1000, \ 3000$ and $\ell_\text{max}= 5000$. The choice $\ell_\text{min} = 2$, where the noise includes the primary CMB, is clearly not optimal as the signal to noise is at most close to one. Raising the filtering scale to $\ell_\text{min} = 1000$, the signal drops slightly (an effect that is more pronounced at low redshift), but the noise drops by an order of magnitude, raising the signal to noise accordingly. Raising the filtering scale further to $\ell_\text{min} = 3000$ again further increases the signal to noise. Clearly, removing as much of the primary CMB as possible through filtering is the choice that will optimize signal to noise. Comparing the dotted curves ($\ell_\text{min} = 3000$) with figure~\ref{fig:sigs}, reducing $\ell_\text{max}=\infty$ to $\ell_\text{max} = 5000$ degrades the signal to noise by about a factor of $10-100$, with the degredation more pronounced in the large-redshift bins. Nevertheless, the signal to noise is still $S/N \sim 10-100$ over a significant range in $L$.


\begin{figure}[htbp]
	\begin{center}
	\includegraphics[width=15.5cm]{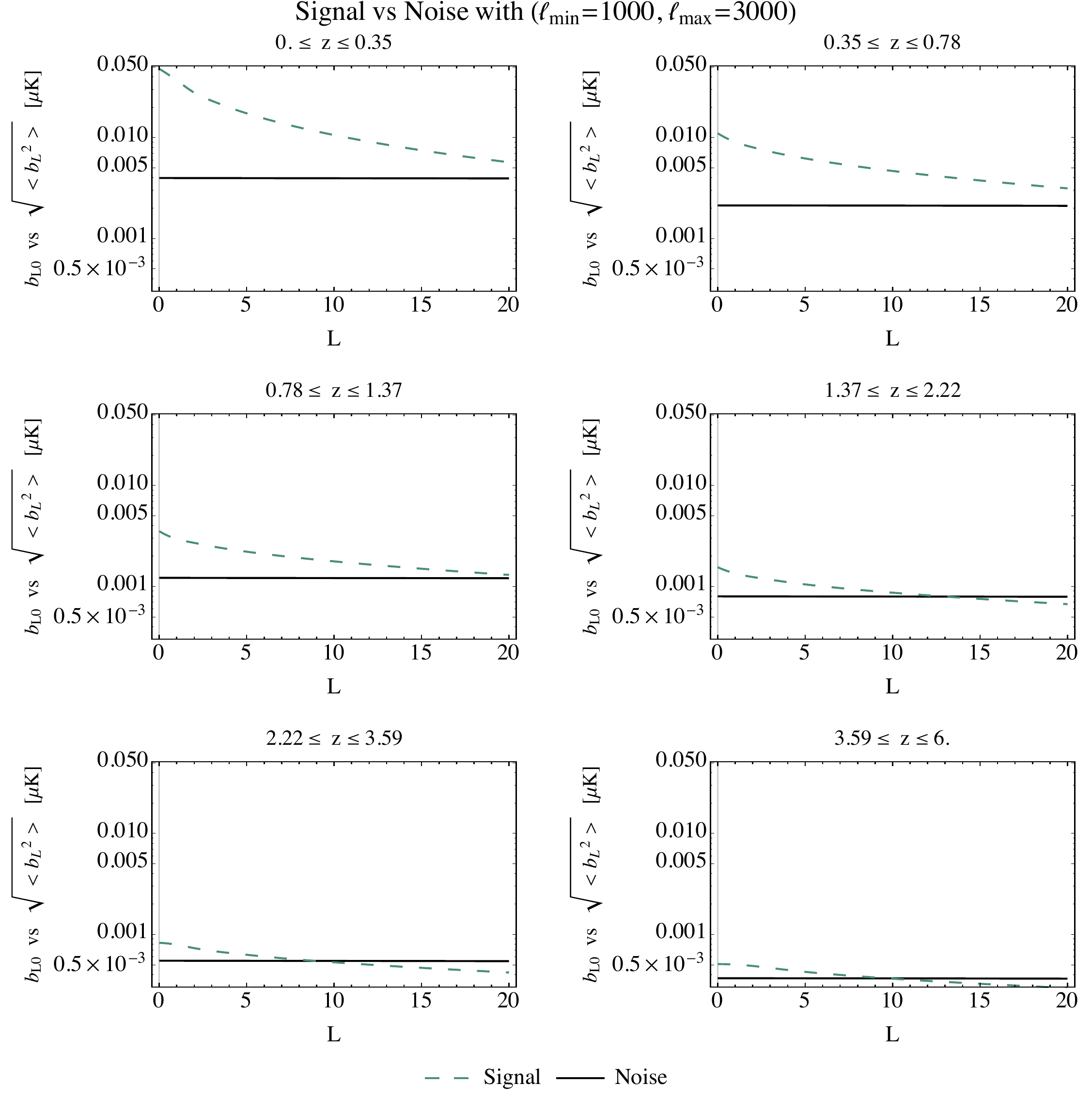}
	\caption{The signal \eqref{eq:signal} (dashed) and noise \eqref{eq:noise} (solid) are shown in six redshift bins, computed with $k_\text{max} = \infty$, $\ell_\text{min}=3000$, and $\ell_\text{max}=3000$.}
	\label{fig:lmax3000}
	\end{center}
\end{figure}

Moving on to the low resolution scenario, in figure~\ref{fig:lmax3000} we show the signal and noise for six redshift bins, choosing $\ell_\text{min} = 1000$ and $\ell_\text{max} = 3000$. Here, it can be seen that the signal-to-noise is significant only for the low redshift bins, and reaches at most $S/N \sim 10$.

With these results, we can map the low and high resolution scenarios to a rough set of experimental requirements. In particular, we consider CMB instrumental noise, foreground residuals, finite resolution, and galaxy shot noise. The treatment of a number of important systematics is beyond the scope of this paper. We assume that the instrumental noise contribution to the CMB temperature is gaussian and uniform on the sky,\footnote{This is clearly incorrect, and spatially varying noise will be an important systematic to assess in the future.} gaussian beams, and that foreground residuals in the cleaned data product used for the analysis can be modelled as a uniform gaussian random field. Under these assumptions, the measured CMB temperature power spectrum is modelled as:
\begin{equation}
C_\ell^{TT} = \left( C_\ell^{TT, \text{kSZ}} + C_\ell^{TT, \text{p}} + N_\ell^{\rm CMB} + F_\ell^{\rm CMB} \right) \exp\left[ \frac{\ell (\ell+1) \theta_{\rm FWHM}^2}{8 \ln 2} \right],
\end{equation}
where $\theta_{\rm FWHM}$ is the full-width at half-maximum of the Gaussian beam in radians. The noise and foreground contributions are 
\begin{equation}
N_\ell^{\rm CMB} = (\sigma_N \ \theta_{\rm FWHM})^2,
\end{equation}
where $\sigma_N$ is the noise per variance in each beam-sized patch, and
\begin{equation}
F_\ell^{\rm CMB} = (\sigma_F \ \theta_{\rm FWHM})^2,
\end{equation}
where $\sigma_F$ is the variance in foreground residuals in each beam-sized patch. 

Assuming that galaxy number density is an unbiased tracer of free electrons,\footnote{We reserve a more complete treatment for future work.} and assuming that redshift bins are far larger than redshift errors, the measurement of the density angular power spectrum is limited by shot noise, yielding:
\begin{equation}
C_{\ell'}^{\delta\delta} (\bar{\chi}_e)=  C_{\ell}^{\text{gg}} (\bar{\chi}_e) + N_{\ell}^{\text{gg}} (\bar{\chi}_e),
\end{equation}
with
\begin{equation}
N_{\ell}^{\text{gg}} (\bar{\chi}_e) = \frac{1}{N_g(\bar{\chi}_e)},
\end{equation}
where $N_g$ is the number of galaxies per square radian in a redshift bin centred on $\bar{\chi}_e$.

We collect the experimental requirements in table~\ref{tab:experiments} for the case of 6 redshift bins. We set the resolution by requiring that the effect of the beam is not dominant, yielding $\theta_{\rm FWHM} = (8 \ln 2 / \ell_\text{max}  (\ell_\text{max} +1))^{1/2}$. The requirements on the CMB instrumental noise and foregrounds are then set by the condition that one measures the CMB power spectrum at a signal to noise of one at $\ell_\text{max}$, e.g. $N_{\ell_\text{max}}^{\rm CMB}, F_{\ell_\text{max}}^{\rm CMB} = C_{\ell_\text{max}}^{TT, \text{kSZ}} + C_{\ell_\text{max}}^{TT, \text{p}}$. Finally, we solve for $N_g(\bar{\chi}_e)$ from the condition that one measures the galaxy-galaxy power spectrum with signal to noise of one, e.g.  $C_{\ell_\text{max}}^{\text{gg}} (\bar{\chi}_e) = N_{\ell_\text{max}}^{\text{gg}} (\bar{\chi}_e)$. 

Our choices for resolution correspond roughly to the range of resolutions considered for Stage 4 CMB experiments~\cite{Abazajian:2016yjj} of $\theta_{\rm FWHM} \simeq 1-3 \ {\rm arc min}$. The requirement on instrumental noise for the two resolution scenarios we consider falls within the target of $\sigma_N \theta_{\rm FWHM} \sim 1 \mu{\rm K} \ {\rm arcmin}^{-1}$~\cite{Abazajian:2016yjj}. It is expected that foreground residuals will not dominate the signal until $\ell \agt 3000$~\cite{Abazajian:2016yjj}, making it unlikely that the high resolution scenario could be achieved without better frequency coverage, as could be attained from space. The forecasted galaxy number density is $N_g \sim 30$ for Euclid~\cite{2011arXiv1110.3193L}, and about  $N_g \sim 130$ for LSST~\cite{LSST09}. This will be sufficient to recover information from the first 2-3 redshift bins (which covers most of the reach of such surveys) in either resolution scenario; a more optimal binning strategy can be constructed for a specific survey. It is unlikely that the higher redshift bins could be accessed with a galaxy survey. For comparison, the galaxy density in the Hubble Deep Field~\cite{1998RvMA...11...83F} is $N_g \sim 500$, and $N_g \sim 1700$ in the Hubble Ultra Deep Field~\cite{Beckwith:2006qi} (which is close to the cosmic variance limit). Since these surveys are fairly complete to high redshift, it is likely that bins 5 and 6 in the high resolution case require more galaxies than there are in the observable Universe. It may be possible to access the high resolution example with new techniques such as intensity mapping. 

\begin{table*}
 \begin{center}
\begin{tabular}{c c c c}
\hline
\hline
$\ell_\text{max}$ & \ $\theta_{\rm FWHM}$ (arcmin) \ & \ $\sigma_N, \ \sigma_F$ ($\mu$K $\theta_{\rm FWHM}^{-1}$) \ & \ $N_g (\bar{\chi}_e)$ \ $({\rm arcmin^{-2}})$ \\
\hline
3000 & 2.7 & 3.7 & 14, 26, 82, 255, 697, 1661 \\
5000 & 0.6 & 3.0 & 45, 62, 155, 500, 1700, 5000 \\
\hline
\hline
 \end{tabular}
 \caption{The experimental characteristics required to access the low-resolution ($\ell_\text{max}=3000$) and high resolution ($\ell_\text{max}= 5000$) scenarios.
 \label{tab:experiments}}
 \end{center}
\end{table*}

In conclusion, it seems that a first detection could in principle be made with the next generation of CMB experiments and galaxy surveys. Based on the rough analysis above, future progress can be made with increases in resolution of CMB experiments, better foreground subtraction using multifrequency information, and new techniques for measuring the angular matter spectrum to high resolution, such as high resolution intensity mapping. A more detailed forecast will be presented in future work.

\section{Discussion and Conclusions} \label{sec:conclusion}

In this paper, we have assessed the ability of kSZ tomography to yield information about the long-wavelength Universe in the cosmic variance limit. The signal of interest is a power asymmetry in the direct cross correlation of the kSZ contribution to the CMB temperature and the electron density binned at various redshifts. We quantify this signal in terms of power multipoles, and compared it with the amplitude of ``accidental" power asymmetry due to the statistically homogeneous component of the kSZ contribution to the CMB temperature (the primary source of cosmic variance in this context). The results are promising in this highly optimistic scenario, yielding a signal to noise greater than unity over a large range of power multiples in a large number of redshift bins. A first forecast indicates that next-generation CMB experiments and galaxy surveys should be able to make a detection at low redshift and large angular scale.

Although we have established that there is in principle a signal to detect, there is a significant amount of work that must be done to assess what can be done in practice. First, a more detailed forecast should be done to determine what is required to reach the necessary threshold in sensitivity, accuracy, and resolution. Next, there are important systematic errors and other potential sources of power asymmetry that should be investigated including, but not limited to: relativistic aberration of the CMB~\cite{Aghanim:2013suk} and large scale structure~\cite{Gibelyou11122012,Yoon:2015lta}, clustering~\cite{1984ApJ...284L...9K}, electron bias in the nonlinear regime (e.g.~\cite{Fox:1997mm}), non-Gaussian aspects of CMB lensing~\cite{Zaldarriaga:2000ud}, redshift space distortions~\cite{Kaiser87}, asymmetric scan strategies (e.g.~\cite{2000ApL&C..37..259D}) and incomplete sky coverage, incomplete LSS surveys, asymmetric beams and point spread functions, and more realistic window functions. Furthermore, the signal may be boosted in the presence of some types of primordial non-Gaussianity~\cite{Adhikari:2015yya}. We leave a more careful investigation of these and other important aspects to future work. 

If it is indeed possible to approach the cosmic variance limited scenario we have described in this paper, what would we stand to learn? Because kSZ tomography is probing different portions of the surface of last scattering than the primary CMB, the constraining power for various early Universe scenarios involving extra sources of inhomogeneity can be extraordinary. For example, in Ref.~\cite{Zhang:2015uta} it was shown that the constraints on parameters in a theory that predicts cosmic bubble collisions could improve by several orders of magnitude comparing kSZ tomography to existing and forecasted constraints from the primary CMB. One might expect constraints on theories of the large-scale CMB anomalies to be similarly impressive. Given a high enough fidelity measurement, it should also be possible to reconstruct the 3D large-scale gravitational potential throughout much of the observable Universe (in analogy with Ref.~\cite{Yadav2005}). Such a reconstruction could be an important tool, for example in future studies of primordial non-Gaussianity. Performing this exercise would also significantly clarify precisely what new information, beyond that encoded in the primary CMB and CMB polarization, there is to gain from kSZ tomography. In any case, we stand to learn a great deal about cosmology in the coming era of precision measurements of CMB secondaries, with the kSZ effect playing a leading role. In this new era, it is important to understand the nuances of the kSZ effect, and to target new observables.

\acknowledgments
We are grateful for helpful discussions with Eiichiro Komatsu and Dragan Huterer.
We also thank Anne-Sylvie Deutsch and Moritz M{\"u}nchmeyer for insightful comments on the draft.
MCJ is supported by the National Science and Engineering Research Council through a Discovery grant. AT acknowledges support from the Vanier Canada Graduate Scholarships program. This research was supported in part by Perimeter Institute for Theoretical Physics. Research at Perimeter Institute is supported by the Government of Canada through the Department of Innovation, Science and Economic Development Canada and by the Province of Ontario through the Ministry of Research, Innovation and Science. Results in this paper were obtained using the Healpix package (https://sourceforge.net/projects/healpix/) and the Cosmicpy package (http://cosmicpy.github.io/).

\appendix
\section{The effective velocity in Fourier space} \label{app:veff}

In this appendix, we derive an expression for the effective line of sight velocity $v_{\rm eff}$ in Fourier space. The Fourier transform of the primordial potential is defined as
\begin{equation}
\Psi_i ({\bf r}) = \int \frac{d^3 k}{(2 \pi)^3} \tilde{\Psi}_i ({\bf k}) e^{i \chi_e {\bf k} \cdot {\bf \hat{n}}_e } e^{i \Delta \chi {\bf k} \cdot {\bf \hat{n}} },  
\end{equation}
where we have explicitly expanded the position ${\bf r} = \chi_e {\bf \hat{n}}_e + \Delta \chi {\bf \hat{n}}$. 

\subsection{Sachs-Wolfe}

Using eq.~\eqref{eq:thetaSW}, the SW contribution to the effective velocity is related to the Fourier components of $\Psi_i$ through 
\begin{align} 
v_{\rm eff, SW}({\bf \hat{n}}_e, \chi_e) =& \frac{3}{4\pi}\left( 2D_\Psi(\chi_\text{dec}) -\frac{3}{2} \right) \int d^2{\bf \hat{n}} \ \Psi_i ({\bf r}) \ \mathcal{P}_1 ({\bf \hat{n}} \cdot  {\bf \hat{n}}_e ) \no \\
=& \frac{3}{4\pi}\left( 2D_\Psi(\chi_\text{dec}) -\frac{3}{2} \right) \int \frac{d^3 k}{(2 \pi)^3} \tilde{\Psi}_i ({\bf k}) e^{i \chi_e {\bf k} \cdot {\bf \hat{n}}_e }   \no \\
 & \times \int d^2{\bf \hat{n}} \  e^{i \Delta \chi {\bf k} \cdot {\bf \hat{n}} }   \ \mathcal{P}_1 ({\bf \hat{n}} \cdot  {\bf \hat{n}}_e ). \label{eq:vsachswolfe1}
\end{align}
We can work on the second integral by expanding the exponential in terms of Legendre polynomials and spherical Bessel functions:
\begin{equation} \label{eq:expidentity}
e^{i \Delta \chi {\bf k} \cdot {\bf \hat{n}} } = \sum_{\ell'} i^{\ell'} (2 \ell' + 1) \ j_{\ell'} (k \Delta \chi) \ \mathcal{P}_\ell ({\bf \hat{k}} \cdot  {\bf \hat{n}} ).
\end{equation}
Substituting and applying the identity,
\begin{equation}\label{eq:doublelegendreidentity}
\int  d^2{\bf\hat{b}} \ \mathcal{P}_{\ell'} ({\bf\hat{a}} \cdot  {\bf\hat{b}}) \ \mathcal{P}_{\ell} ({\bf\hat{b}} \cdot {\bf\hat{c}}) = \frac{4 \pi}{2 \ell + 1} \mathcal{P}_{\ell} ({\bf\hat{a}} \cdot  {\bf\hat{c}}) \delta_{\ell \ell'} ,
\end{equation}
results in
\be \label{eq:legendreexp}
\int d^2{\bf \hat{n}} \  e^{i \Delta \chi {\bf k} \cdot {\bf \hat{n}} }   \ \mathcal{P}_1 ({\bf \hat{n}} \cdot  {\bf \hat{n}}_e ) = 4 \pi i \ j_{1} (k \Delta \chi) \ \mathcal{P}_{1} ({\bf \hat{k}} \cdot  {\bf \hat{n}}_e ) .
\ee
Putting this result back into eq.~\eqref{eq:vsachswolfe1} gives
\ba \label{eq:vSW}
v_{\rm eff, SW}({\bf \hat{n}}_e, \chi_e) &=& 3 i \left( 2D_\Psi(\chi_\text{dec}) -\frac{3}{2} \right) \int \frac{d^3 k}{(2 \pi)^3} \tilde{\Psi}_i ({\bf k}) j_{1} (k \Delta \chi_\text{dec}) \mathcal{P}_{1} ({\bf \hat{k}} \cdot  {\bf \hat{n}}_e )e^{i \chi_e {\bf k} \cdot {\bf \hat{n}}_e }.
\ea
We therefore see that $v_{\rm eff, SW}$ is simply a convolution of the potential field evaluated on the intersection of the electron's past light cone and the time of decoupling.

\subsection{Doppler}

Using eq.~\eqref{eq:thetaDopp}, the Doppler contribution to $v_{\rm eff}$ is
\ba \label{eq:vDoppler1}
v_{\rm eff, Doppler}({\bf \hat{n}}_e, \chi_e) &=& \frac{3}{4\pi} D_v (\chi_\text{dec})  \int d^2{\bf \hat{n}} \ ( {\bf \hat{n}} \cdot {\bf \nabla} \Psi_i ({\bf r}_{\rm dec})) \ \mathcal{P}_1 ({\bf \hat{n}} \cdot  {\bf \hat{n}}_e ) \\
&-& \frac{3}{4\pi} D_v (z_{\rm e})  \int d^2{\bf \hat{n}} \ ( {\bf \hat{n}} \cdot {\bf \nabla} \Psi_i ({\bf r}_{\rm e})) \ \mathcal{P}_1 ({\bf \hat{n}} \cdot  {\bf \hat{n}}_e ).
\ea
We'll start with the first integral, then do the second. Going to Fourier space and expanding the exponent in Legendre polynomials gives
\begin{align}
 & \frac{3}{4\pi} D_v (\chi_\text{dec}) \int \frac{d^3 k}{(2 \pi)^3} i k \tilde{\Psi}_i ({\bf k}) \ e^{i \chi_e {\bf k} \cdot {\bf \hat{n}}_e } \int  d^2{\bf \hat{n}} \ \mathcal{P}_1 ({\bf \hat{n}} \cdot {\bf \hat{k}} ) \ e^{i \Delta \chi_\text{dec} {\bf k} \cdot {\bf \hat{n}} } \ \mathcal{P}_1 ({\bf \hat{n}} \cdot  {\bf \hat{n}}_e ) \no \\
= & \frac{3}{4\pi} D_v (\chi_\text{dec}) \int \frac{d^3 k}{(2 \pi)^3} i k \tilde{\Psi}_i ({\bf k}) \ e^{i \chi_e {\bf k} \cdot {\bf \hat{n}}_e } \sum_{\ell'} i^{\ell'} (2 \ell' + 1) j_{\ell'} (k \Delta \chi_\text{dec} )   
 \no\\
 & \times \int d^2{\bf \hat{n}} \ \mathcal{P}_1 ({\bf \hat{n}} \cdot {\bf \hat{k}} ) \ \mathcal{P}_{\ell'} ({\bf \hat{n}} \cdot {\bf \hat{k}} ) \ \mathcal{P}_1 ({\bf \hat{n}} \cdot  {\bf \hat{n}}_e ).
\end{align}

The integral over three Legendre polynomials can be evaluated by expanding in spherical harmonics and using the Wigner 3j symbols. The result is
\be
\int d^2{\bf \hat{n}} \ \mathcal{P}_1 ({\bf \hat{n}} \cdot {\bf \hat{k}} ) \ \mathcal{P}_{\ell'} ({\bf \hat{n}} \cdot {\bf \hat{k}} ) \ \mathcal{P}_1 ({\bf \hat{n}} \cdot  {\bf \hat{n}}_e ) =  \frac{4 \pi}{3} \mathcal{P}_1 ({\bf \hat{k}} \cdot  {\bf \hat{n}}_e ) \delta_{\ell' 0} + \frac{8 \pi}{15} \mathcal{P}_1 ({\bf \hat{k}} \cdot  {\bf \hat{n}}_e ) \delta_{\ell' 2} ,
\ee
and upon substitution, we obtain for the first integral:
\be
i D_v (\chi_\text{dec}) \int \frac{d^3 k}{(2 \pi)^3} \ k \ \tilde{\Psi}_i ({\bf k}) \ \left[ (j_0(k \Delta \chi_\text{dec} ) -2 j_{2} (k \Delta \chi_\text{dec} ) \right] \ \mathcal{P}_1 ({\bf \hat{k}} \cdot  {\bf \hat{n}}_e )  \ e^{i \chi_e {\bf k} \cdot {\bf \hat{n}}_e }.
 \ee
Moving to the second integral in eq.~\eqref{eq:vDoppler1}, we have
\be
- \frac{3}{4\pi} D_v (\chi_{e})   \int \frac{d^3 k}{(2 \pi)^3} i k \tilde{\Psi}_i ({\bf k}) e^{i \chi_e {\bf k} \cdot {\bf \hat{n}}_e } \ \int d^2{\bf \hat{n}} \ \mathcal{P}_1 ( {\bf \hat{n}} \cdot {\bf \hat{k}}) \ \mathcal{P}_1 ({\bf \hat{n}} \cdot  {\bf \hat{n}}_e ).
\ee
The integral over angles can be evaluated using the identity eq.~\eqref{eq:doublelegendreidentity} to obtain
\be
- i D_v (\chi_{e})   \int \frac{d^3 k}{(2 \pi)^3}  \ k\ \tilde{\Psi}_i ({\bf k})  \ \mathcal{P}_1 ({\bf \hat{k}} \cdot  {\bf \hat{n}}_e ) \ e^{i \chi_e {\bf k} \cdot {\bf \hat{n}}_e }.
\ee
Assembling the various pieces, the Doppler contribution becomes:
\begin{align}
	v_{\rm eff, Doppler}({\bf \hat{n}}_e, \chi_e) = &  i \int \frac{d^3 k}{(2 \pi)^3} \ k \tilde{\Psi}_i ({\bf k}) \mathcal{P}_1 ({\bf \hat{k}} \cdot  {\bf \hat{n}}_e )  \ e^{i \chi_e {\bf k} \cdot {\bf \hat{n}}_e } \no \\
	& \times \left[ D_v (\chi_\text{dec}) j_{0} (k \Delta \chi_\text{dec} ) -2 D_v (\chi_\text{dec}) j_{2} (k \Delta \chi_\text{dec} ) - D_v (\chi_{e}) \right].
	\label{eq:veffDoppler}
\end{align}
Note that the last term in square brackets contains no spherical Bessel function. Therefore, the Doppler component receives contributions from all scales, unlike the Sachs-Wolfe term.

\subsection{Integrated Sachs-Wolfe}

Using eq.~\eqref{eqn:ISW}, the ISW contribution to the effective velocity is
\be
v_{\rm eff, ISW}({\bf \hat{n}}_e, \chi_e) = \frac{3}{4 \pi} \int d^2 {\bf \hat{n}} \left( 2 \int_{a_{\rm dec}}^{a_e} \frac{dD_\Psi}{da}\Psi_i({\bf r}(a)) da \right)\mathcal{P}_1 ({\bf \hat{n}} \cdot  {\bf \hat{n}}_e ).
\ee
Going to Fourier space, 
\be
v_{\rm eff, ISW}({\bf \hat{n}}_e, \chi_e) = \frac{3}{2 \pi} \int_{a_{\rm dec}}^{a_e} da \frac{dD_\Psi}{da} \int \frac{d^3 k}{(2 \pi)^3} \tilde{\Psi}_i ({\bf k}) e^{i \chi_e {\bf k} \cdot {\bf \hat{n}}_e } \ \int d^2 {\bf \hat{n}} \ e^{i \Delta \chi (a) {\bf k} \cdot {\bf \hat{n}} }   \mathcal{P}_1 ({\bf \hat{n}} \cdot  {\bf \hat{n}}_e ),
\ee
and applying the identity eq.~\eqref{eq:legendreexp}, we obtain
\begin{equation} \label{eq:vISW}
v_{\rm eff, ISW}({\bf \hat{n}}_e, \chi_e) = 6 i \int \frac{d^3 k}{(2 \pi)^3} \tilde{\Psi}_i ({\bf k})  \left[ \int_{a_{\rm dec}}^{a_e} da \frac{dD_\Psi}{da}  \ j_{1} (k \Delta \chi (a)) \right] \mathcal{P}_{1} ({\bf \hat{k}} \cdot  {\bf \hat{n}}_e )  e^{i \chi_e {\bf k} \cdot {\bf \hat{n}}_e } .
\end{equation}
Just as for the SW contribution, the ISW contribution is mainly sensitive to potential fluctuations on large scales. 

\subsection{Effective velocity}
We can now assemble eqs.~\eqref{eq:vSW}, \eqref{eq:veffDoppler}, \eqref{eq:vISW} into an expression for the total effective velocity. Before doing so, it must be noted that the expression eq.~\eqref{eq:veffDoppler} for the Doppler kernel is only valid in the small-$k$ limit, so the linear growth with $k$ eventually gets cutoff. To fix this, we can incorporate the transfer function $T(k)$ by simply replacing $\tilde{\Psi}_i ({\bf k}) \rightarrow T(k) \tilde{\Psi}_i ({\bf k})$. We will employ the BBKS fitting function:
\begin{equation}
T(k) = \frac{\ln \left[ 1 + 0.171 x \right]}{0.171 x} \left[ 1+ 0.284 x + (1.18 x)^2 + (0.399 x)^3 + (0.49 x)^4 \right]^{-0.25},
\end{equation}
where $x = k / k_{\rm eq}$ with $k_{\rm eq} = a_{\rm eq} H(a_{\rm eq}) = \sqrt{2/a_{\rm eq}} H_0 \simeq 82.5 H_0$. Putting together all three components, \eqref{eq:vSW}, \eqref{eq:veffDoppler}, \eqref{eq:vISW}, gives the expression for the effective velocity:
\begin{equation}
v_{\rm eff}({\bf \hat{n}}_e, \chi_e) = i \int \frac{d^3 k}{(2 \pi)^3} \ T(k) \tilde{\Psi}_i ({\bf k}) \ \mathcal{K}^v (k, \chi_e) \ \mathcal{P}_{1} ({\bf \hat{k}} \cdot  {\bf \hat{n}}_e ) \ e^{i \chi_e {\bf k} \cdot {\bf \hat{n}}_e },
\end{equation}
where $\mathcal{K}^v=\mathcal{K}_{\rm D} +\mathcal{K}_{\rm SW} +\mathcal{K}_{\rm ISW}  $ is the full Fourier kernel with each component given by,
\ba
\mathcal{K}_{\rm D} (k, \chi_e) &\equiv& k D_v (\chi_\text{dec}) j_{0} (k \Delta \chi_\text{dec} ) - 2 k D_v (\chi_\text{dec}) j_{2} (k \Delta \chi_\text{dec} ) - k D_v (\chi_e),  \\ 
\mathcal{K}_{\rm SW} (k, \chi_e) &\equiv& 3\left( 2D_\Psi(\chi_\text{dec}) -\frac{3}{2} \right) j_{1} (k \Delta \chi_\text{dec}), \\ 
\mathcal{K}_{\rm ISW} (k, \chi_e) &\equiv& 6  \int_{a_{\rm dec}}^{a_e} da \frac{dD_\Psi}{da}  \ j_{1} (k \Delta \chi (a)).
\ea

\section{A pure gradient is pure gauge} \label{app:gradientisgauge}

In this appendix we explicitly demonstrate that a pure gradient in the Newtonian potential $\Psi$ (or more generally, the curvature perturbation in an arbitrary gauge) can be removed through a special conformal transformation. More generally, we can remove the gradient of the Newtonian potential at a point, which we take to be the origin of Cartesian coordinates. The Newtonian potential appears in a conformal factor in front of the spatial metric:
\begin{equation}
ds_3^2 = (1 - 2 \Psi({\bf x})) \delta_{ij} dx^i dx^j .
\end{equation}
Performing a special conformal transformation
\begin{equation}\label{eq:sctrans}
x^i = \frac{{x'}^{i} - b^i {x'}_i {x'}^i}{ 1 - 2 b_i {x'}^i + \left(b_i b^i \right) \left( {x'}_i {x'}^i \right)} ,
\end{equation}
takes the spatial metric to
\begin{equation}\label{eq:sctransmetric}
\delta_{ij} dx^i dx^j = \frac{\delta_{ij}}{\left[ 1 - 2 b_i {x'}^i + \left(b_i b^i \right) \left( {x'}_i {x'}^i \right) \right]^2} d{x'}^i d{x'}^j ,
\end{equation}
where $b_i$ are free constants.

If we imagine there was a pure gradient in the Newtonian potential,
\begin{equation}
\Psi ({\bf x})= A_i x^i ,
\end{equation}
we can write
\begin{align}
(1 - 2 \Psi ({\bf x}) ) \delta_{ij} dx^i dx^j = & \left(1 - 2  A_i \frac{{x'}^{i} - b^i {x'}_i {x'}^i}{ 1 - 2 b_i {x'}^i + \left(b_i b^i \right) \left( {x'}_i {x'}^i \right)} \right)  \no \\
& \times\frac{\delta_{ij}d{x'}^i d{x'}^j}{\left[ 1 - 2 b_i {x'}^i + \left(b_i b^i \right) \left( {x'}_i {x'}^i \right) \right]^2} .
\end{align}
For $b_i \ll 1$, expanding to first order, we have:
\begin{equation}
(1 - 2 \Psi ({\bf x}) ) \delta_{ij} dx^i dx^j \simeq \left(1 - 2 (A_i - 2 b_i)x'^i + \mathcal{O} (b^2) \right) \delta_{ij} dx'^i dx'^j .
\end{equation}
Therefore, with the choice,
\begin{equation}
b_i = A^i / 2,
\end{equation}
we have $\Psi ({\bf x}') = 0$, and therefore no gradient in the primed coordinate system. Note also that the choice for $b_i$ justifies neglecting the terms of higher order in $b$, at least in linear perturbation theory.

More generally, we could imagine performing a Taylor series expansion of $\Psi ({\bf x})$ about a point, taken here to be ${\bf x} = 0$:
\begin{equation}
\Psi ({\bf x}) \simeq \Psi (0) + \partial_i \Psi (0) x^i + \ldots
\end{equation}
If we perform a special conformal transformation with
\begin{equation}
b_i = \partial_i \Psi (0) / 2,
\end{equation}
then in the primed coordinates we have
\begin{equation}
\partial_i \Psi (0) x^i = \frac{1}{2} - \frac{1 - 3 \partial_i \Psi(0) \ {x'}^i + \left(\partial_i \Psi(0) \ \partial^i \Psi(0) \right) \left( {x'}_i {x'}^i \right)/4 + (\partial^i \Psi(0) \ {x'}^i)^2}{ 2 \left[1 - \partial_i \Psi(0) \  {x'}^i + \left(\partial_i \Psi(0) \ \partial^i \Psi(0) \right) \left( {x'}_i {x'}^i \right)/4 \right]^3}.
\end{equation}
To lowest order in $\partial_i \Psi(0)$, this is
\begin{equation}
\partial_i \Psi (0) x^i \simeq (\partial_i \Psi(0) \ {x'}^i)^2 + \left(\partial_i \Psi (0) \  \partial^i \Psi (0)\right) \left( {x'}_i {x'}^i \right)/4.
\end{equation}
Therefore, in the primed coordinate system, the Taylor series expansion of $\Psi ({\bf x}')$ is
\begin{equation}
\Psi ({\bf x}') \simeq \Psi (0) + \mathcal{O}(x'^2) + \ldots
\end{equation}
with no linear term as advertised. Therefore, the special conformal transformation can be used to eliminate the derivative of $\Psi$ at a point. This comes at the price of altering the higher-order terms in the Taylor series expansion.

\section{Cancellation of the kernel contributions as $k \rightarrow 0$ for $\Lambda$CDM without radiation} \label{sec:appB}

In this Appendix we show the exact cancellation of the three contributions to the kernel of the effective velocity given in \eqref{eq:kernel}-\eqref{eq:kernel3} for the largest scales in a universe with only matter and $\Lambda$. In this case, $y=a/a_\text{eq} \rightarrow \infty$, so we can approximate $D_\Psi(a)$ using
\be \label{eq:dpsi}
D_\Psi(a)\equiv \frac{\Psi_{\rm SH}(a)}{\Psi_{\rm
  SH,i}}=\frac{9}{10} \left[ \frac{5}{2}\Omega_m \frac{E(a)}{a} G(a)\right],
\ee
where $G(a) \equiv \int_0^a da^{'} \left[ E(a^{'})a^{'}\right]^{-3}$ and $E(a)=\sqrt{\Omega_ma^{-3}+\Omega_\Lambda}$. Further, the distance along the electron's past light cone to redshift $z=1/a-1$, normalized by $H_0$ is given by
\be
	\Delta\chi(a)=-\int_{a_e}^a da^{'} \frac{1}{E(a^{'}){a^{'}}^2} .
\ee

Let's begin with the simple Sachs-Wolfe term. Expanding to linear order in $k$, we obtain
\begin{align}
	\mathcal{K}_\text{SW} = & 3\left( 2D_\Psi(\chi_\text{dec}) -\frac{3}{2} \right) j_{1} (k \Delta \chi_\text{dec})\no\\
	=& 3\left( 2D_\Psi(\chi_\text{dec}) -\frac{3}{2} \right) \frac{k \Delta \chi_\text{dec}}{3} + \mathcal{O}(k^3)\no\\ \label{eq:ksw}
	=& \left( 2D_\Psi(\chi_\text{dec}) -\frac{3}{2} \right) k \Delta \chi_\text{dec} + \mathcal{O}(k^3) .
\end{align}

Expanding the Bessel functions, the Doppler piece will only have two terms at linear order:
\be
	\mathcal{K}_\text{D}=-kD_v(a_e)+kD_v(a_\text{dec})+\mathcal{O}(k^3)
\ee
We can simplify $D_v$ as follows
\begin{align}
	D_v(a) =& \frac{2a^2H(a)}{\Omega_m} \frac{y}{4+3y} \left[D_\Psi (a)+\frac{dD_\Psi (a)}{d\ln a}\right] \no\\
	=& \frac{2a^2E(a)}{\Omega_m} \frac{1}{3}\left[D_\Psi (a)+a\frac{dD_\Psi (a)}{da}\right]\no \\
	=& \frac{2a^2E(a)}{3\Omega_m} \left[-\frac{3\Omega_m}{2E^2(a)a^3}\left( D_\Psi (a) -\frac{3}{2} \right) \right]\no\\
	=& -\frac{1}{aE(a)}\left( D_\Psi (a) -\frac{3}{2} \right) .
\end{align}
This allows us to write the Doppler piece of the kernel as
\begin{align}
	\mathcal{K}_\text{D} =& \frac{k}{a_eE(a_e)}\left( D_\Psi (a_e) -\frac{3}{2} \right) -\frac{k}{ a_\text{dec} E(  a_\text{dec} )}\left( D_\Psi (  a_\text{dec} ) -\frac{3}{2} \right) +\mathcal{O}(k^3) \no\\ 
	=& \frac{3k}{2}\left(    \frac{1}{ a_\text{dec} E(  a_\text{dec} )} - \frac{1}{a_eE(a_e)} \right) + \frac{9k}{4} \Omega_m  \left( \frac{G(a_e)}{a_e^2} -  \frac{G( a_\text{dec} )}{ a_\text{dec}^2 } \right)+\mathcal{O}(k^3).\label{eq:kdop}
\end{align}

The ISW term can be shown to exactly cancel the above two contributions. We start by expanding $j_{1} (k \Delta \chi (a)) \sim (k \Delta \chi (a))/3 $ and integrating by parts:
\begin{align}
	\mathcal{K}_\text{ISW} =& 6  \int_{a_{\rm dec}}^{a_e} da \frac{dD_\Psi (a)}{da}  \ j_{1} (k \Delta \chi (a))  \no \\
	=& \left[ 6D_\Psi (a) \frac{k \Delta \chi (a)}{3} \right]^{a=a_e}_{a=a_\text{dec}} -6\int_{a_{\rm dec}}^{a_e} da \frac{k D_\Psi (a) }{3}\frac{d \Delta \chi (a)  }{da} 	+\mathcal{O}(k^3) \no \\
	=& -2D_\Psi ( a_\text{dec} ) k \Delta \chi_\text{dec} + 2k \int_{a_{\rm dec}}^{a_e} da \frac{D_\Psi (a) }{a^2 E(a)} +\mathcal{O}(k^3) \no\\ 
	=& -2D_\Psi ( a_\text{dec} ) k \Delta \chi_\text{dec} + \frac{9k}{2}\Omega_m \int_{a_{\rm dec}}^{a_e} da \frac{G (a) }{a^3} +\mathcal{O}(k^3) .\label{eq:iswsub}
\end{align}
The integral in the second term becomes
\begin{align}
	\int_{a_{\rm dec}}^{a_e} da \frac{G (a) }{a^3} =& \left[ \frac{-G(a)}{2a^2} \right]^{a=a_e}_{a=a_\text{dec}} + \int_{a_{\rm dec}}^{a_e} \frac{da}{2a^5E^3(a)} \no\\
	 =& -\frac{1}{2}\left( \frac{G(a_e)}{a_e^2} -  \frac{G( a_\text{dec} )}{ a_\text{dec}^2 } \right) + \frac{1}{3\Omega_m} \int_{a_{\rm dec}}^{a_e} \frac{da}{a} \frac{d}{da}\left[ \frac{1}{E(a)}\right] \\
	  =& -\frac{1}{2}\left( \frac{G(a_e)}{a_e^2} -  \frac{G( a_\text{dec} )}{ a_\text{dec}^2 } \right) +\frac{1}{3\Omega_m}\left( \left[ \frac{1}{aE(a)} \right]^{a=a_e}_{a=a_\text{dec}} + \int_{a_{\rm dec}}^{a_e} \frac{da}{a^2E(a)} \right)\no\\
	  =& -\frac{1}{2}\left( \frac{G(a_e)}{a_e^2} -  \frac{G( a_\text{dec} )}{ a_\text{dec}^2 } \right) -\frac{1}{3\Omega_m} \left( \frac{1}{ a_\text{dec} E(  a_\text{dec} )} - \frac{1}{a_eE(a_e)}   + \Delta\chi_\text{dec}  \right).\no
\end{align}
Inserting this into \eqref{eq:iswsub} gives the final result for the ISW contribution
\begin{align} \label{eq:kisw}
	\mathcal{K}_\text{ISW} =& -2D_\Psi ( a_\text{dec} ) k \Delta \chi_\text{dec}  -\frac{9k}{4}\Omega_m\left( \frac{G(a_e)}{a_e^2} -  \frac{G( a_\text{dec} )}{ a_\text{dec}^2 } \right) \no \\
	& -\frac{3k}{2}\left( \frac{1}{ a_\text{dec} E(  a_\text{dec} )} - \frac{1}{a_eE(a_e)}  + \Delta\chi_\text{dec}  \right) +\mathcal{O}(k^3)
\end{align}
It is now clear that adding equations \eqref{eq:ksw}, \eqref{eq:kdop} and \eqref{eq:kisw} gives
\be
	\mathcal{K}_\text{SW} + \mathcal{K}_\text{D} +\mathcal{K}_\text{ISW} = 0 +\mathcal{O}(k^3).
\ee

When radiation is included, this cancellation still holds as shown numerically in figure~\ref{fig:kernel}.

\section{Random Gaussian fields}\label{sec:randomfields}

Following Ref.~\cite{0067-0049-137-1-1}, we generate realizations of the primordial potential $\Psi({\bf x})$ in a four-step process. Given a spatial grid of size $L^3$ with $N^3$ positions ${\bf x} ({\bf m})= L {\bf m } / N$, labeled by the integer triplet ${\bf m}$ with components $m_i \in [0,N)$:
\begin{enumerate}
\item Define a field $\xi ({\bf m})$ that lives on the grid. Draw $\xi ({\bf m})$ at each ${\bf m}$ from an independent Gaussian pdf with variance $N^3$.
\item Fourier transform to get 
\begin{equation}
\xi ({\bf \kappa}) = N^{-3} \sum_{\bf m} \exp \left[ - \frac{2 \pi i}{N} {\bf \kappa} \cdot {\bf m} \right] ,
\end{equation}
where $\kappa \equiv k L / (2 \pi)$ is the dimensionless wavenumber.
\item Multiply $\xi ({\bf \kappa})$ by
\begin{equation}
F(k) \equiv \left[ \left( \frac{2 \pi}{L} \right)^3 P_\Psi (k) \right]^{1/2},
\end{equation}
where $P_\Psi (k)$ is the Gaussian primordial power spectrum of $\Lambda$CDM. Here, we use parameters for the amplitude and spectral index consistent with Planck~\cite{Planck:2015} ($A_s = 2.2 \times 10^{-9}$, $n_s = 0.96$).
\item Inverse Fourier transform to obtain a random field with the correct correlation properties in real and Fourier space:
\begin{equation}
\Psi_i ({\bf m}) = \sum_{{\bf \kappa}} F(k) \xi({\kappa}) \exp \left[ \frac{2 \pi i}{N} {\bf \kappa} \cdot {\bf m} \right].
\end{equation}

\end{enumerate}

\bibliographystyle{JHEP}
\bibliography{refs}

\end{document}